\documentclass[manuscript]{aastex}

\usepackage{natbib, aas_macros}
\usepackage{amsmath}
\usepackage{color}

\usepackage{ulem}

\definecolor{MyDarkBlue}{rgb}{0,0.08,0.45}
\definecolor{MyDarkRed}{rgb}{0.8,0.1,0.08}
\definecolor{Red}{rgb}{1.0,0.0,0.2}
\definecolor{Blue}{rgb}{0,0.08,0.95}
\definecolor{LightGrey}{rgb}{0.7,0.7,0.7}

\begin{document}
\title{Efficient small-scale dynamo in solar convection zone}
\author{H. Hotta$^{1}$, M. Rempel$^1$, and T. Yokoyama$^2$}
\affil{
$^1$High Altitude Observatory, National Center for Atmospheric Research,
Boulder, CO, USA\\
$^2$Department of Earth and Planetary Science, University of Tokyo,
7-3-1 Hongo, Bunkyo-ku, Tokyo 113-0033, Japan
}
\email{ hotta@ucar.edu}
\begin{abstract}
\end{abstract}
We investigate small-scale dynamo action in the solar convection zone
through a series of high resolution MHD simulations in a local Cartesian
domain with  $1\,R_{\odot}$ (solar radius) of horizontal extent and a radial 
extent from $0.715$ to $0.96\,R_{\odot}$.
The dependence of the solution on resolution and diffusivity is studied.
For a grid spacing of less than 350 km, the root mean square
magnetic field strength near the base of the convection zone reaches
95\% of the equipartition field strength (i.e. magnetic and kinetic
energy are comparable). 
For these solutions the Lorentz force feedback on the convection
velocity is found to be significant. The velocity near the base of the convection
zone is reduced to 50\% of the hydrodynamic one. In spite of a significant
decrease of the convection velocity, the reduction in the enthalpy flux
is relatively small, since the magnetic field also suppresses the horizontal
mixing of the entropy between up- and downflow regions. This effect
increases the amplitude of the entropy perturbation and makes convective
energy transport more efficient. We discuss potential implications of these
results for solar global convection and dynamo simulations.

\keywords{Sun: interior --- Sun: dynamo --- Stars: interiors}
\clearpage
\section{Introduction}
Turbulent thermal convection fills the solar convection zone due to
its superadiabatic stratification. Thermal convection under the influence of
rotation leads to angular momentum transport and maintains large-scale
mean flows, such as the differential rotation and the meridional flow. These flows 
in combination with helical turbulence play a crucial role for maintaining the Sun's 
global magnetic field through a large-scale dynamo, which has been widely studied 
through meanfield and 3D approaches
\citep{1955ApJ...122..293P,1969AN....291...49S,1969AN....291..271S,1980opp..bookR....K,2003A&ARv..11..287O,2005LRSP....2....2C,2005LRSP....2....1M}.
In addition, also non-helical turbulent thermal convection itself
has the ability to maintain a turbulent magnetic field (small-scale
dynamo e.g. \cite{1950RSPSA.201..405B}).
%{\bf e.g. Batchelor Proc. R. Soc. London A 201, 405 (1950)}.
Small-scale dynamos maintain magnetic fields on scales comparable to 
or smaller than the energy carrying scale of turbulence
\citep[see, e.g.,][section 5]{2005PhR...417....1B} and are sometimes also called
``fluctuation dynamos'' \citep[see also][]{2012SSRv..169..123B}.
Numerical simulations of a small-scale dynamo were presented first by 
\cite{1981PhRvL..47.1060M} for non-stratified forced turbulence, followed by high 
resolution small-scale dynamo simulations in the context
of galaxies, stars, the Sun as well as idealized forcing turbulence 
\citep[e.g.][]{1996JFM...306..325B,1996ApJ...464..690H,1999ApJ...515L..39C,2004PhRvE..70a6308H,2004PhRvL..92e4502S}.
Small-scale dynamos are excited when the magnetic Reynolds
number exceeds a critical value, typically near $100$, which is found to be dependent on the
magnetic Prandtl number $\mathrm{Pm}=\nu/\eta$
\citep{2004PhRvE..70a6308H,2004PhRvL..92e4502S,2005ApJ...625L.115S,2007PhRvL..98t8501I,2007NJPh....9..300S}.
The critical magnetic Reynolds number is found to be larger by a factor of a few for the $\mathrm{P_m} \ll 1$ regime that
is most relevant for stellar and planetary dynamos.
Compared to large-scale dynamos, small-scale dynamos have in their kinematic phase small growth time scales
that are determined by the eddy turn-over time scale near the resistive scale ($\mathrm{Pm}\ll 1$) or viscous scale 
($\mathrm{Pm}\gg 1$), i.e. the growth rate is strongly dependend on the magnetic (and viscous) diffusivity. 
While the onset and kinematic growth of small-scale dynamos has been studied in great detail, the
non-linear saturation phase in particular for the $\mathrm{Pm}\ll 1$ regime has been addressed only very recently
\citep{2011AN....332...51B,2014ApJ...791...12B}.

The non-linear saturation regime of a small-scale dynamo is most relevant for a theoretical explanation of 
the observed mixed polarity field in the solar photosphere outside
active regions, the so-called ``quiet sun''. Small-scale dynamo simulations that include more realistic effects, such as partial ionization and radiation
in the uppermost few Mm of the convection zone were presented recently by \cite{2007A&A...465L..43V,0004-637X-714-2-1606}.
\cite{2014ApJ...789..132R} found that explaining the observed level of quiet Sun magnetic field requires solutions 
that reach a magnetic field strength close to equipartition only a few Mm beneath the photosphere. 
Small-scale magnetic field with such strength has potentially a
significant influence on convective dynamics.

The above mentioned photospheric dynamo simulations cover only the uppermost few Mm of the solar convection zone and
typically use open bottom boundary conditions to ``mimic'' the influence from the deeper convection zone. Due to the use of open
boundary conditions these simulations cannot determine the saturation field strength of a small-scale dynamo in the photosphere
without assumptions about the magnetic field strength and structure in the deeper convection zone. We present here small
scale dynamo simulations that are very similar to those presented by \cite{2014ApJ...789..132R}, but focus on the lower 
convection zone ranging from $0.715$ to $0.96\,R_{\odot}$.

The purpose of this study is twofold: Firstly we want to investigate 
whether the assumption about the magnetic field strength in the convection zone made in \cite{2014ApJ...789..132R} in order to 
explain the observed quiet Sun field strength are realistic. Secondly we
want to investigate how a strong (about equipartition) small-scale
magnetic field influences convection and convective energy transport
throughout the convection zone.

In global dynamo simulations small-scale and large-scale dynamo action is present at the same time, however most global
dynamo simulations have insufficient resolution to properly address the small-scale dynamo: The sun has a circumference
of 4400 Mm at the surface, which is large compared to the typical scales of the
convective structures and the local pressure scale height ($\sim
60\ \mathrm{Mm}$ at the base of the convection zone). Most investigations of
the global solar dynamo have a grid spacing larger than
$4\ \mathrm{Mm}$ \citep{2002ApJ...570..865B,2013arXiv1310.8417A,2014ApJ...789...35F}. This
resolution is not sufficient to resolve the inertial range of the turbulence,
which is important for the small-scale dynamo. While the small-scale dynamo is not completely absent, its
efficiency and saturation field strength are likely significantly underestimated.
 
Recently \cite{2014ApJ...786...24H} (hereafter Paper I) studied a small-scale dynamo in a global setup without
rotation (in order to rule out large-scale dynamo action by construction) and
achieved a grid spacing of $1\ \mathrm{Mm}$ and found rather efficient
small-scale dynamo action throughout the convection zone. Magnetic field of $0.15B_\mathrm{eq}-0.25B_\mathrm{eq}$ is
maintained by the small-scale dynamo, where $B_\mathrm{eq}=\sqrt{4\pi\varrho} v_\mathrm{rms}$ is the
equipartition magnetic field strength. Although they could see
feedback from the generated magnetic field on the velocity, the
effect is not significant for that resolution.

The purpose of this study is to increase the resolution further by
limiting the size of the calculation domain and to investigate the
efficiency of the small-scale dynamo and its feedback on convection using a grid
spacing as small as $350\ \mathrm{km}$ (a global convection simulation with this resolution would require a grid
of approximately $600(r)\times 6000(\theta) \times 12000(\phi)$). In addition we use (similar to \cite{2014ApJ...789..132R}) a
LES (Large Eddy Simulation) approach, which drastically reduces
diffusivities on the resolved scales of the simulations.
Our approach implies that the effective magnetic Prandtl number
is close to unity and we do not carry out a parameter survey for
the magnetic Prandtl number in this study.
Our focus is on evaluating the potential influence of an efficient small-scale dynamo on convective dynamics, which is currently 
not resolvable in global models. This is quite different from studies
of global dynamos in the Sun and in rapidly rotating systems
\citep{2004ApJ...614.1073B,2010ApJ...715L.133G,2011ApJ...731...69B,2014ApJ...789...35F,1999GeoJI.138..393C,2011GApFD.105..234B,2013Icar..225..185Y,2014A&A...564A..78S,2014arXiv1407.3187Y},
where a significant fraction of
the magnetic energy resides on the largest scales.

\section{Model}
We solve the three-dimensional magnetohydrodynamic equations with the
reduced speed of sound technique \citep[RSST:][]{2012A&A...539A..30H} in
Cartesian geometry ($x$,$y$,$z$). In our definition, $x$-direction is
the vertical direction. The equations are expressed as
\begin{eqnarray}
 \frac{\partial \rho_1}{\partial t} &=&
  -\frac{1}{\xi^2}\nabla\cdot(\rho_0{\bf v}), \\
\rho_0\frac{\partial {\bf v}}{\partial t} &=&
-\rho_0({\bf v}\cdot\nabla){\bf v} -
\nabla\left(p_1+\frac{B^2}{8\pi}\right) + \nabla\cdot\left(\frac{{\bf
       BB}}{4\pi}\right) - \rho_1g{\bf e_x},\\
\frac{\partial {\bf B}}{\partial t}&=&\nabla\times({\bf v}\times{\bf B}),\\
\rho_0T_0\frac{\partial s_1}{\partial t} &=& -\rho_0T_0({\bf
 v}\cdot\nabla)s_1
+\frac{d}{dx}\left(\kappa_r\rho_0c_p\frac{dT_0}{dx}\right) +
\Gamma,\\
p_1 &=& \left(\frac{\partial p}{\partial \rho}\right)_s\rho_1
+\left(\frac{\partial p}{\partial s}\right)_\rho s_1. \label{eos}
\end{eqnarray}
The subscripts 0 and 1 denote that background and time-dependent
perturbed values.
Except for the geometry, the setting is similar to that presented in
Paper I. While a limited horizontal extent of the calculation
domain is required in order to increase the resolution, we also intend to avoid
boundary effect from the side boundaries. Thus we use a periodic boundary and
Cartesian geometry. One disadvantage of this setting is that it leads to the
incorrect energy flux in the upper part of the domain. The solar energy flux at the base of the
convection zone ($r=0.715R_\odot$, where $R_\odot$ is the solar radius),
is about $1.21\times10^{11}\ \mathrm{erg\ s^{-1}\ cm^{-2}}$. This is about
twice as large as the flux found in the photosphere due to the expansion factor of
$r^2$. We should keep in mind that the obtained convective velocity will
be larger by less than factor of $\sqrt{2}$ in the upper part of the
domain.
The scaling law of $v^{2-3}\propto F$ is found in some studies even
in rapid rotation situation
\citep{2010SSRv..152..565C,2014A&A...564A..78S,2015ApJ...798...51H},
which is not far from mixing length theory, which predicts the
dependence of $v^3\propto
F$ \citep[e.g.][]{2004suin.book.....S}.
The background
stratification, $\rho_0$, $p_0$, and $T_0$, is obtained by solving the
hydrostatic equation in the adiabatic atmosphere. The gravitational
acceleration $g$ and the radiative diffusivity $\kappa_\mathrm{r}$ are
adopted from the Model S
\citep{1996Sci...272.1286C}. The obtained stratification is almost same
as the Model S (see Fig. 1 of Paper I). The same
form for the surface cooling $\Gamma$ is adopted as in Paper I. The
thickness of the
cooling layer is two pressure scale heights at the top boundary, which is 18.8
Mm. We use a realistic equation of state, which
include the partial ionization effects using OPAL repository
\citep{1996ApJ...456..902R} is used in eq. (\ref{eos}), the plasma is
almost completely ionized in the calculation domain of this study.
Thus the equation of state is almost the same as that for an ideal gas.
The factor for the RSST is set for keeping the
reduced adiabatic
speed of sound $\hat{c_s}=c_s/\xi$ constant as
\begin{eqnarray}
 \xi(x) = \xi_0\frac{c_\mathrm{s}(x)}{c_\mathrm{s}(x_\mathrm{min})},
\end{eqnarray}
where $c_s(x)=\sqrt{(\partial p/\partial \rho)_s}$ and
$x_\mathrm{min}$ the original adiabatic speed of sound and the location of the bottom boundary. We use
$\xi_0=150$
and the reduced speed of sound $\hat{c_\mathrm{s}}$ is $1.5\
\mathrm{km\ s^{-1}}$. Considering
the validity of the RSST, i.e.
$v_\mathrm{rms}/{\hat{c_\mathrm{s}}}<0.7$, we can properly treat the
convection of $v_\mathrm{rms}<1.05\ \mathrm{km s^{-1}}$, where
$v_\mathrm{rms}$ is root mean square (RMS) velocity.
In addition to increasing
the effective resolution, we also adopt
a less diffusive artificial
viscosity suggested in \cite{2014ApJ...789..132R}. We note that this
modification has a significant effect on resolving the small-scale
magnetic field and leads to significantly larger kinematic growth rates, i.e. more
efficient small-scale dynamos. 

Our calculation domain extends
$(0.715R_\odot,0,0)<(x,y,z)<(0.96R_\odot,R_\odot,R_\odot)$.
The periodic boundary condition is adopted for the $y$- and
$z$-directions ($x$ is vertical). The symmetric boundary condition is adopted for density
and the entropy perturbation
% ($\partial \rho_1/\partial x = \partial s_1/\partial x=0$) 
at the top and bottom boundaries. Impenetrable and stress-free
boundary condition are adopted for the velocity 
($v_x=\partial v_y/\partial x=\partial v_z/\partial x=0$). 
At the top boundary, the
horizontal magnetic field is zero ($B_y=B_z=0$).
At the bottom boundary a symmetric boundary
condition is adopted for the horizontal magnetic field.
%($\partial B_y/\partial x=\partial B_z/\partial x=0$).
Then the vertical
magnetic field $B_x$ is calculated to satisfy the $\nabla\cdot{\bf B}=0$
constraint at top and bottom boundaries. In the calculation domain the
divergence free condition for the magnetic field is maintained with the
method suggested by \cite{2002JCoPh.175..645D}.\par
Our investigation is focused on the solar convection zone. Since we investigate
the contribution from a small-scale dynamo alone, we do not include rotation.
In this setup the stratification and the luminosity are well determined by a solar standard model 
\citep{1996Sci...272.1286C}. The free parameters in this study are domain extent and resolution
(which translates into diffusivities since we use an LES (Large Eddy Simulation) approach). The typical
scale of stratified convection is determined by the local pressure scale height, which
reaches a value of $\sim$ 60 Mm at the base of the convection zone. Our horizontal extent
of 700 Mm is large enough to expect at best a weak influence. The vertical extent of our
domain from $0.715$ to $0.96\,R_{\odot}$ was chosen to capture most of the convection zone,
but leaves out scales near the top, which are too small to be resolved. Since we focus here on
small-scale dynamo simulations it is not sufficient to just resolve the scale of convection, we need
to be able to resolve also a significant fraction of the turbulent energy cascade. The effective magnetic
and fluid Reynolds numbers are tied to the grid resolution in our study due to the use of an LES approach. 
Estimating the fluid and magnetic Reynolds numbers is a
difficult task \citep{2014ApJ...789..132R}. We present one simulation with an explicit magnetic diffusivity
for which we find a value of $\mathrm{Rm}$ in the range from $300$ to
$1200$ (see following paragraph). In our highest resolution LES
simulation the small-scale dynamo has an about $100$ times larger growth rate, which indicates a substantially
(at least 2 orders of magnitude) larger "effective" magnetic Reynolds number in those setups. (Note that 
scaling arguments suggest for the growth rate of a small-scale dynamo $\gamma \sim \sqrt{\mathrm{Rm}}$, which would imply
an effective $\mathrm{Rm}$ a factor of $10^4$ larger, however, \citet{0004-637X-714-2-1606,2014ApJ...789..132R} found
more a  $\gamma \sim \mathrm{Rm}$ relationship for the resolutions currently accessible by numerical simulations.)
\par
We carry out four hydrodynamic cases and corresponding four MHD cases.
The difference is mainly in the resolution (see Table
\ref{param}). The characters ``H'' and ``M'' denote hydrodynamic and
MHD. Cases H256D and M256D include explicit thermal
diffusivity, kinetic viscosity and
magnetic diffusivity in order to compare the result with previous solar
global dynamo calculations. The thermal diffusivity($\kappa$), kinetic
viscosity($\nu$) and the magnetic
diffusivity($\eta$) are added as:
\begin{eqnarray}
 \rho_0\frac{\partial {\bf v}}{\partial t} &=& [...] -\nabla\cdot
  {\bf D},\\
 \frac{\partial {\bf B}}{\partial t} &=& [...]
  -\nabla\times(\eta\nabla\times {\bf B}),\\
\rho_0 T_0\frac{\partial s_1}{\partial t} &=& [...] +
 \nabla\cdot \left(\kappa\rho_0T_0 \nabla s_1\right)\nonumber\\
 &&+2\rho_0\nu\left[e_{ij}e_{ij}-\frac{1}{3}(\nabla\cdot{\bf v})^2\right]+
 \frac{\eta}{4\pi}(\nabla\times{\bf B})^2,
\end{eqnarray}
where ${\bf D}$ is the viscous stress tensor defined as
\begin{eqnarray}
 D_{ij} = -2\rho_0\nu\left[e_{ij}-\frac{1}{3}(\nabla\cdot{\bf
		      v})\delta_{ij}\right],
\end{eqnarray}
and $e_{ij}$ is the strain rate tensor. The distribution of $\nu$ and
$\eta$ are similar to \cite{2014ApJ...789...35F}. At the top boundary,
the value is $1\times10^{12}\ \mathrm{cm^2\ s^{-1}}$ and these decrease
with depth following a $1/\sqrt{\rho_0}$ profile. This setting is common
in ASH (Anelastic Spherical Harmonic) calculation
\citep{2000ApJ...532..593M,2004ApJ...614.1073B}.
In this study we adopt the same value for the thermal diffusivity
as the other values ($\kappa=1\times10^{12}\ \mathrm{cm^2\ s^{-1}}$).
We use the definition of the magnetic Reynolds number
$\mathrm{Re}=v_\mathrm{rms}d_\mathrm{c}/\eta$
\citep[e.g.][]{2012A&A...546A..19G}, where $d_\mathrm{c}$ is the
thickness of the convection zone.
The estimated magnetic Reynolds number is 336 and 1250 at the top and
bottom of the calculation domain, respectively.
$v_\mathrm{rms}$ is obtained from the result.
This is the comparable
value to the recent high resolution calculations
\citep{2010ApJ...711..424B,2011ApJ...731...69B,2012A&A...546A..19G,2014ApJ...789...35F,2014Icar..241..148J,2014arXiv1407.3187Y}
%\citep{2011A&A...533A.108S,2012A&A...546A..19G,2012ApJ...746...51H,2013Icar..225..185Y,2014Icar..241..148J}.
All calculations started from a hydrodynamic case and were evolved for
100 days. 
Then, weak random magnetic field ($B_z$) with amplitude of 100 G is
added to initiate a small-scale dynamo. The imposed magnetic field is uniform in the $z-$direction and random
in the $x-y$ plane.
The name of cases becomes ``M''
correspondingly. In the highest resolution case (H2048 and M2048), the
grid spacing is smaller than 350 km.

\section{Results}
\subsection{Structure of velocity and magnetic field}
\begin{figure}[htbp]
 \centering
 \includegraphics[width=16cm]{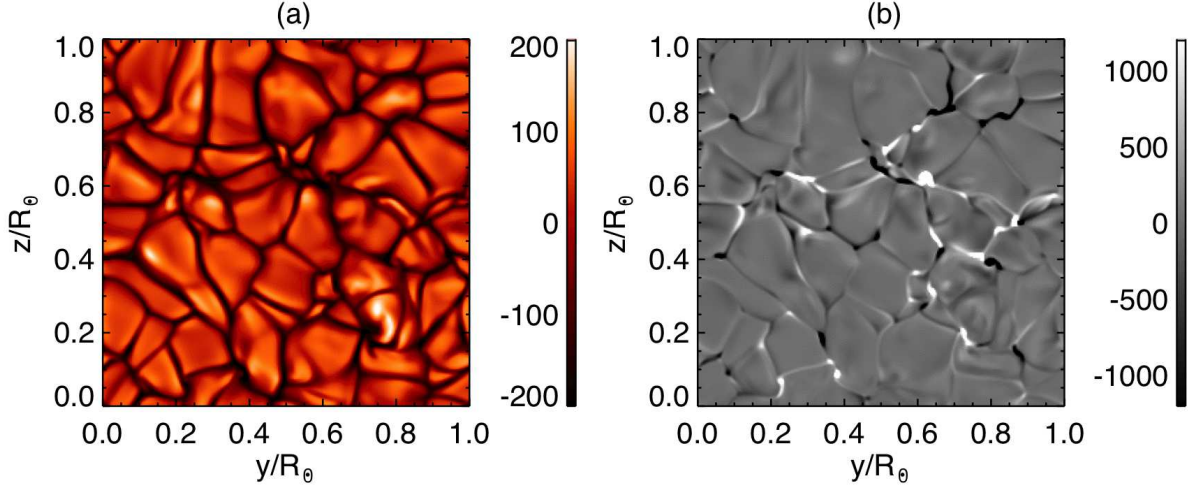}
 \caption{ Contours of (a) vertical velocity $v_x$ in the unit of
 $\mathrm{m\ s^{-1}}$ and (b) vertical
 magnetic field $B_x$ in the unit of $\mathrm{G}$ are shown at
 $r=0.95R_\odot$ for case M256D.
 \label{M256Dvxbx}}
\end{figure}

\begin{figure}[htbp]
 \centering
 \includegraphics[width=10cm]{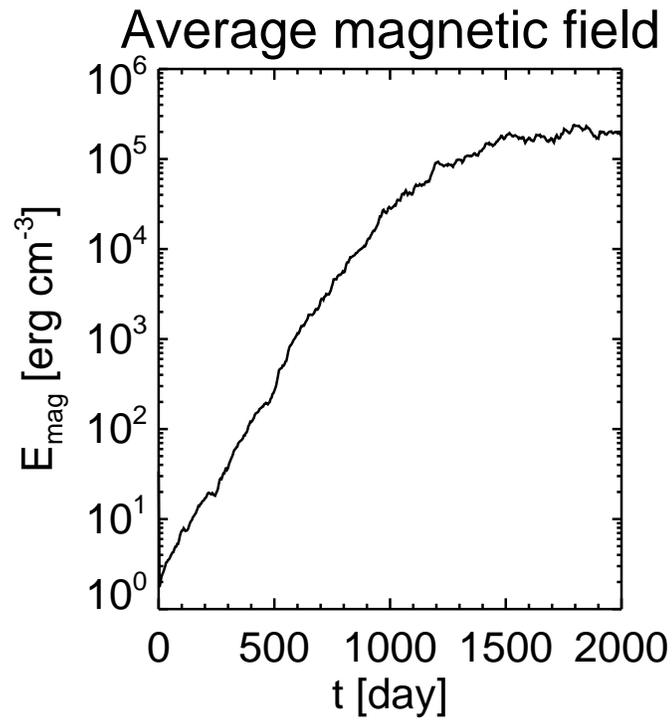}
 \caption{ Temporal evolution of magnetic energy averaged over the whole
 calculation domain in case M256D.
 \label{M256Dmagene}}
\end{figure}

\begin{figure}[htbp]
 \centering
 \includegraphics[width=15cm]{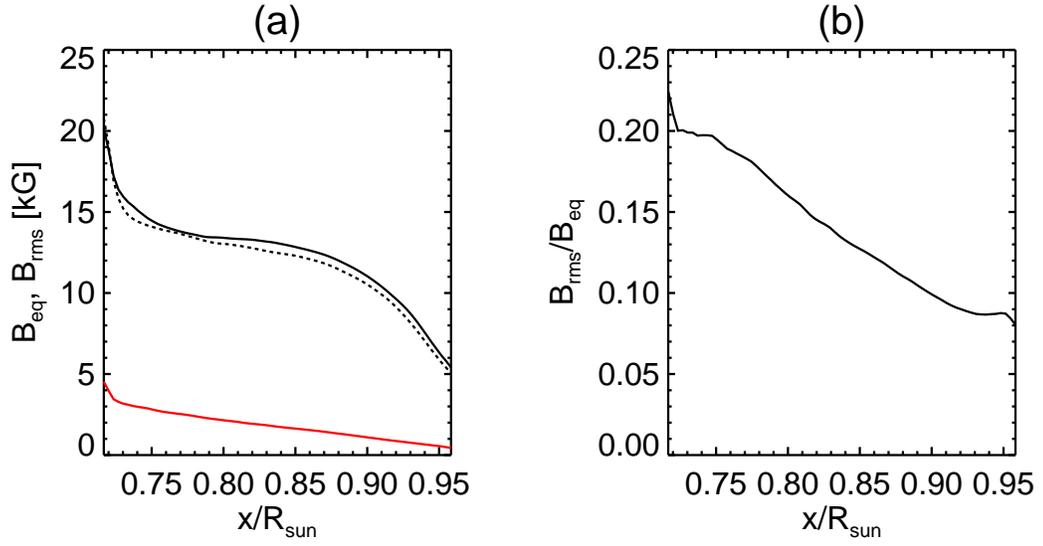}
 \caption{ (a) Black lines shows the equipartition magnetic field to the
 kinetic energy ($B_\mathrm{eq}=\sqrt{4\pi\rho_0}v_\mathrm{rms}$). Solid
 and dotted lines shows results in MHD (M256D) and hydrodynamic (H256D)
 cases. Red line shows the RMS magnetic field. (b) The ratio of
 $B_\mathrm{rms}$ to $B_\mathrm{eq}$ in case M256D.
 \label{M256Dbeq}}
\end{figure}

We start our discussion with cases H256D and M256D. These mimic the
currently available global dynamo calculations. The grid spacing is 2700
km and the explicit diffusivities are included. 
We note that almost isotoropic grid spacing is used in this paper,
i.e, $\Delta x\sim\Delta y=\Delta z$.
We calculate case M256D for 2000 days in order to cover its long time
scale to the saturated phase.
Figs. \ref{M256Dvxbx}a
and b
show the contour of vertical velocity $v_x$ and vertical magnetic field
$B_x$ at $r=0.95R_\odot$, respectively. As is often seen in global dynamo
calculations \cite[e.g.][]{2004ApJ...614.1073B}, broad upflows are
surrounded by thin downflows. The
vertical magnetic field is concentrated in the downflow region.
Fig. \ref{M256Dmagene} shows the temporal evolution of magnetic energy
averaged over the computational domain in case M256D. The e-folding time
scale for magnetic energy during the kinematic phase is 112 days. This is longer 
than the convective time
scale at the base of the convection zone, which is
$H_\rho/v_\mathrm{rms}\sim 10\ \mathrm{day}$, where $H_\rho$ and
$v_\mathrm{rms}$ are the density scale height and the RMS velocity,
respectively. We note that when we double the diffusivities, i.e., 
$\kappa=\nu=\eta=2\times10^{12}\ \mathrm{cm^2\ s^{-1}}$, we do not find
an excited small-scale dynamo.
Fig. \ref{M256Dbeq}a shows the distribution of the equipartition magnetic
field strength
$B_\mathrm{eq}=\sqrt{4\pi\rho_0}v_\mathrm{rms}$ (black lines) and the
RMS magnetic field $B_\mathrm{rms}$ (red line). The dotted black line
shows the result of the hydrodynamic case (H256D). The RMS magnetic
field is about 3000 G and 500 G around the bottom and top boundaries,
respectively. The small difference of solid and dotted black lines shows that
the influence of the Lorentz feedback on the convection is very small in
this case. The dynamo in case M256D maintains
$0.08B_\mathrm{eq}-0.20B_\mathrm{eq}$ magnetic field.
The result indicates that the small-scale dynamo in
currently available global dynamo calculation is not efficient and that the
generated magnetic fields have only a negligible effect on the convective
velocity. 
These findings concern only to the contribution from small-scale field.
The large-scale field of a dynamo in rapidly rotating systems can have a 
significantly effect on the turbulence in some studies
\citep[e.g.][]{1999GeoJI.138..393C,2014arXiv1407.3187Y}.
In the following investigation we decrease the grid
spacing to values of less than 350 km, which
currently cannot be reached on the global dynamo calculation.\par
\begin{figure}[htbp]
 \centering
 \includegraphics[width=15cm]{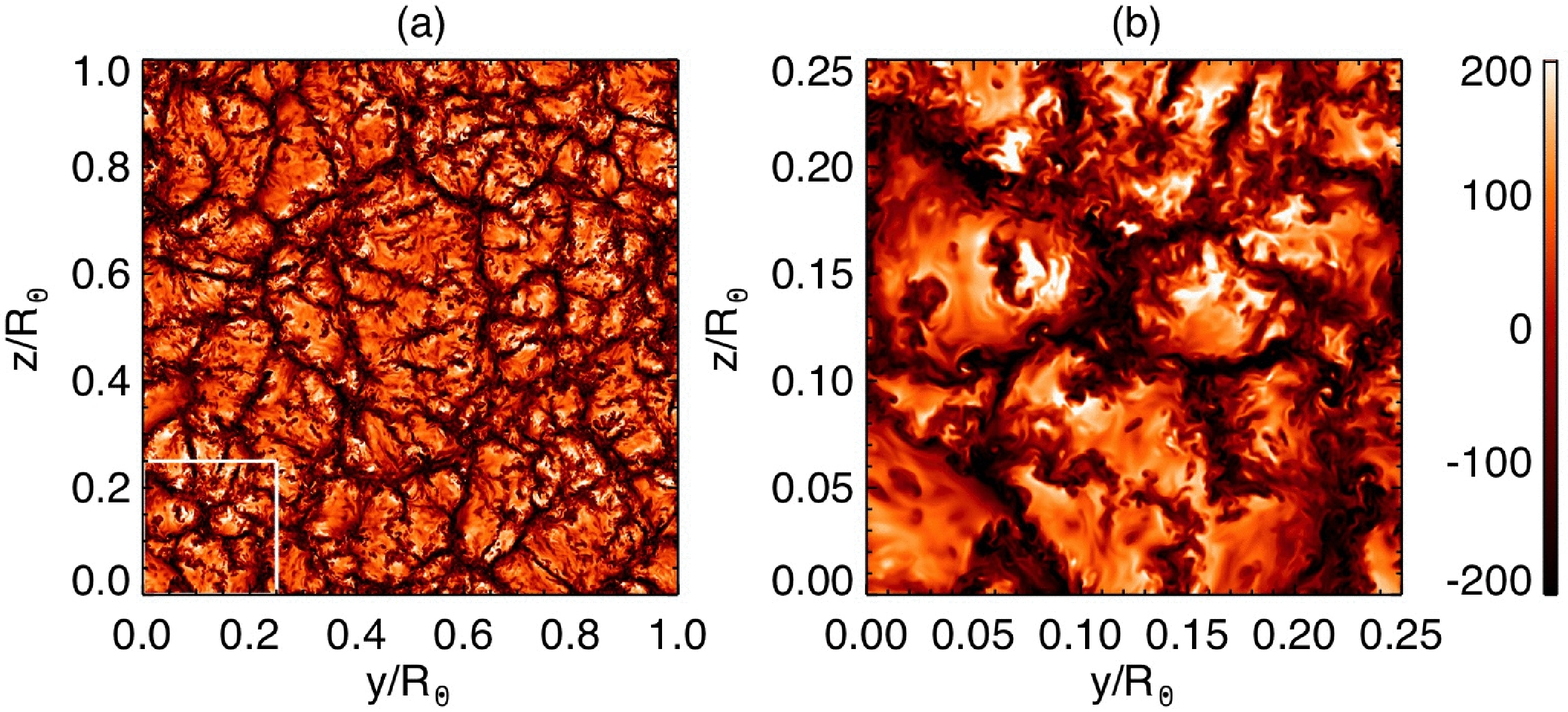}
 \includegraphics[width=15cm]{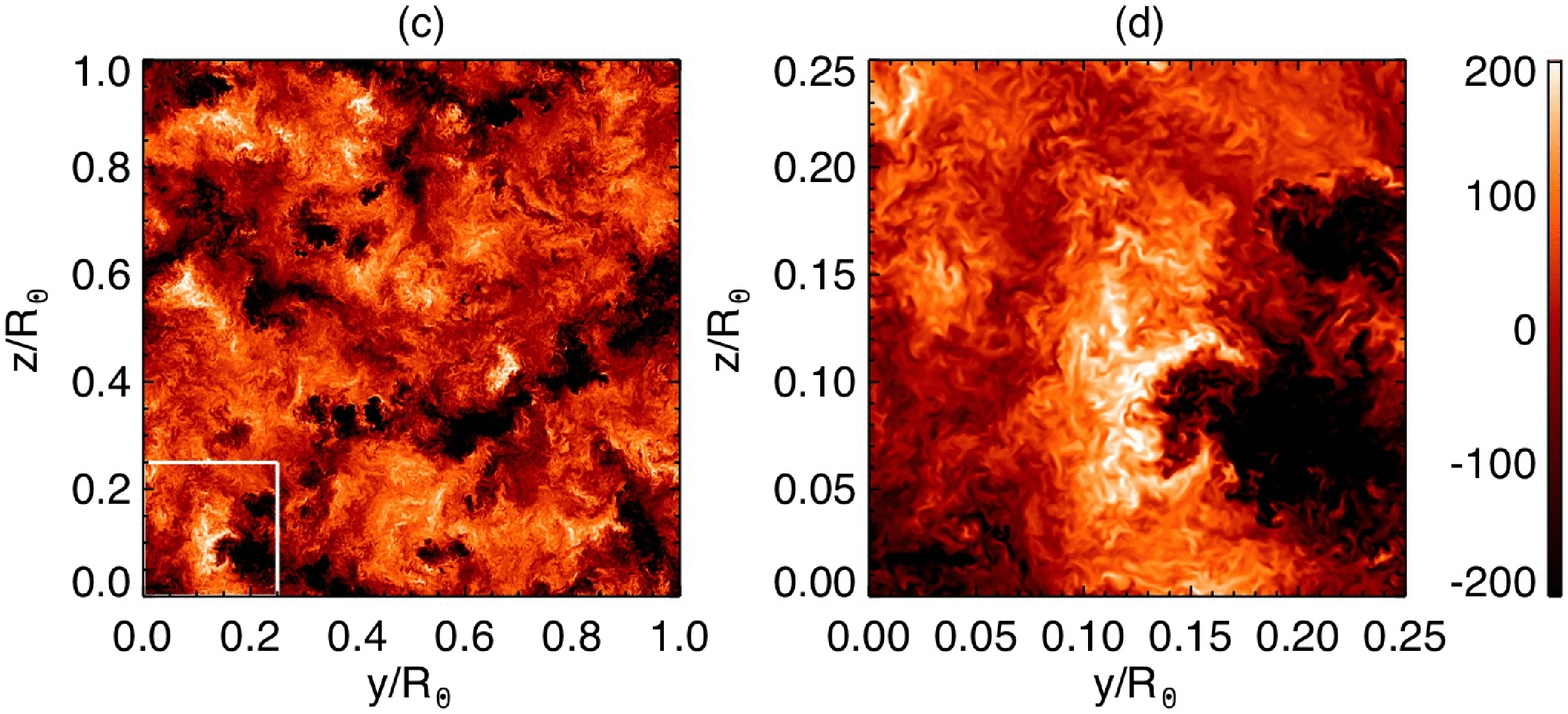}
 \caption{ The distribution of vertical velocity $v_x$ in case H2048 at
 $r=0.95R_\odot$ (panels a and b) and $r=0.8R_\odot$ (panels c and d) in
 the unit of $\mathrm{m\ s^{-1}}$.
 Panels b and d show the zoom in image of panels a and c, respectively.
 The white rectangles indicate the region for panels b and d.
 \label{H2048vx}}
\end{figure}
\begin{figure}[htbp]
 \centering
 \includegraphics[width=16cm]{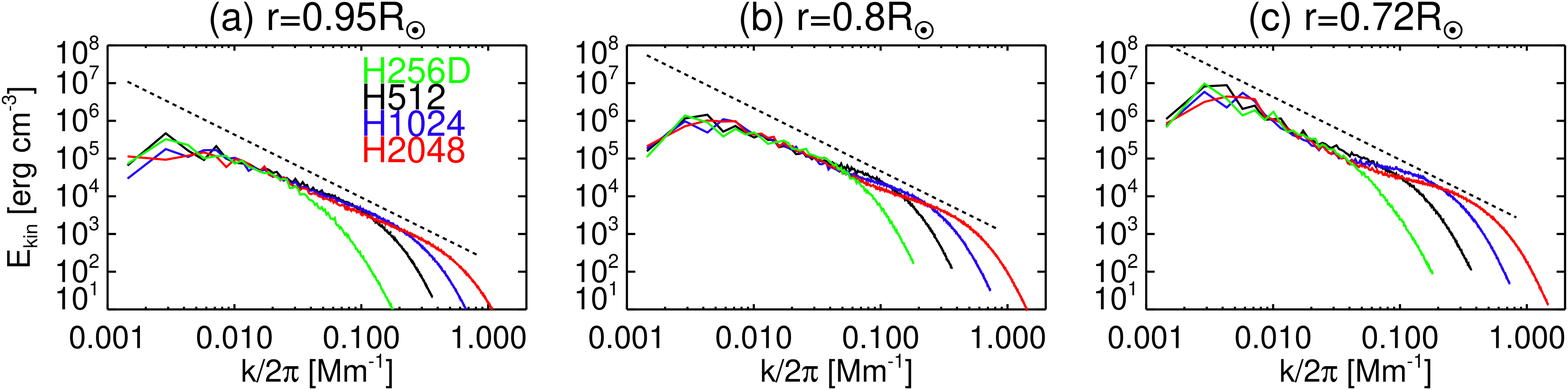}
 \caption{
 Spectra of kinetic energy at (a) $r=0.95R_\odot$, (b) $r=0.8R_\odot$,
 and (c) $r=0.72R_\odot$. Green, black, blue, red lines show the results
 of H256D, H512, H1024, and H2048, respectively. The dotted lines show a
 Kolmogorov slope of -5/3.
 \label{comp_f_hydro}}
\end{figure}

All simulations  with higher resolution do not use any explicit diffusivities. 
The investigation on hydrodynamic cases
(H512, H1024, and H2048) is done for the time interval 80-100 days. Fig. \ref{H2048vx}
shows the distribution of the vertical velocity at $r=0.95R_\odot$
(panels a and b) and $r=0.80R_\odot$ (panels c and d). Compared with
case M256D (Fig. \ref{M256Dvxbx}a), convection cells are distorted by
small-scale turbulence (Fig. \ref{H2048vx}b). Following the increase
of pressure scale height, the typical convection cell size becomes large in the
deeper layer. In addition, small-scale turbulence exists much even in the upflow region
(Fig. \ref{H2048vx}d). 
Figs. \ref{comp_f_hydro}a, b, and c show the spectra of kinetic energy
at $r=0.95R_\odot$, $0.8R_\odot$, and $0.72R_\odot$, respectively for
cases H256D, H512, H1024, and H2048. The dotted line indicates a Kolmogorov slope of
power-law index $-5/3$. In the upper layer ($r=0.95R_\odot$), the spectra of
kinetic energy obey the power-low distribution with the index of
-5/3 in cases without explicit diffusivities. In the deeper region, especially in $r=0.72R_\odot$, there is
a depression around
$k/2\pi=2\times10^{-1}\ \mathrm{Mm^{-1}}$ which is largest in
case H2048 (red line in Fig. \ref{comp_f_hydro}c).\par
\begin{figure}[htbp]
 \centering
 \includegraphics[width=16cm]{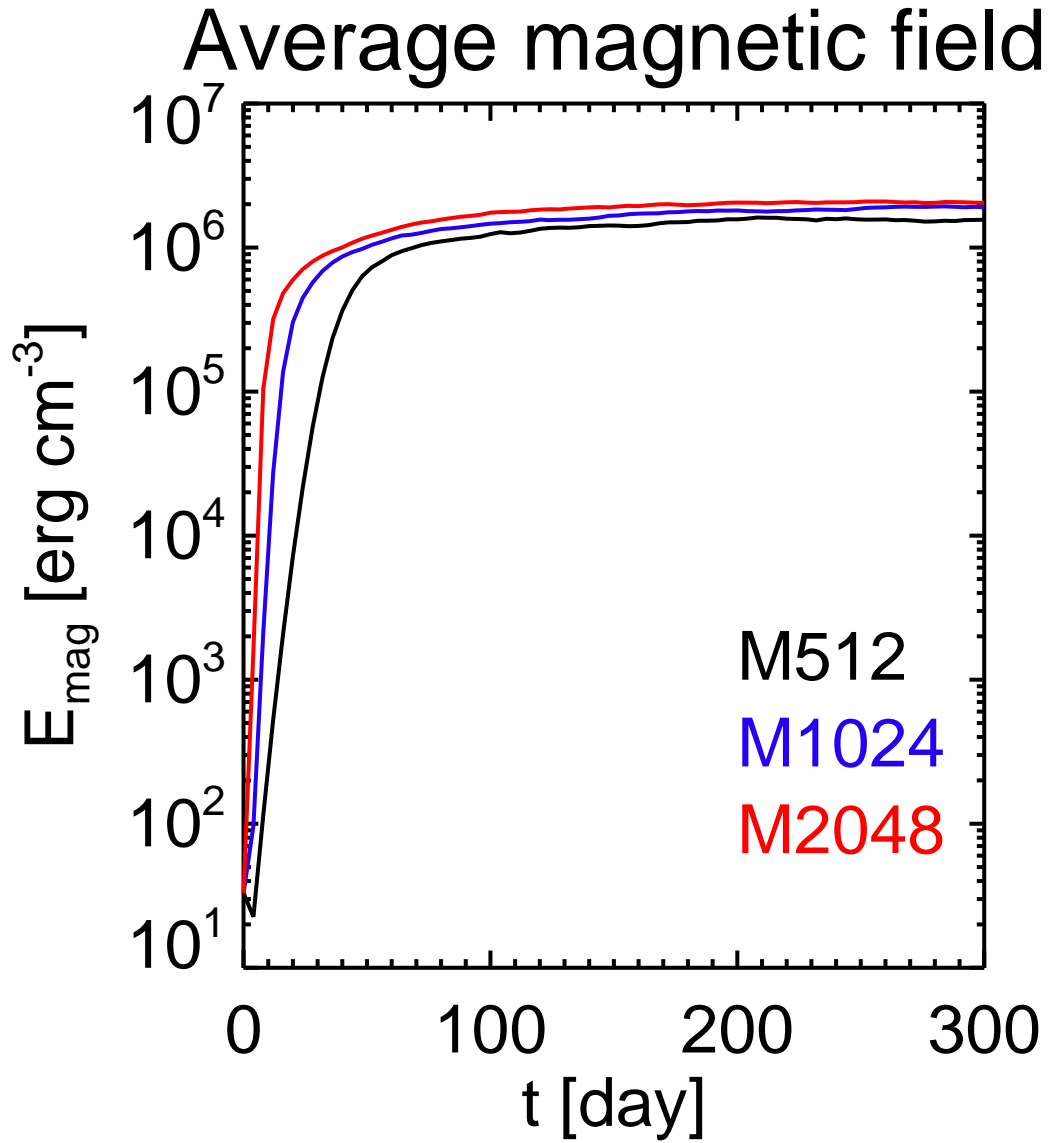}
 \caption{
 Temporal evolution of magnetic energy. The black, blue, and red lines
 show the results in cases M512, M1024, and M2048.
 \label{comp_magene}}
\end{figure}
\begin{figure}[htbp]
 \centering
 \includegraphics[width=16cm]{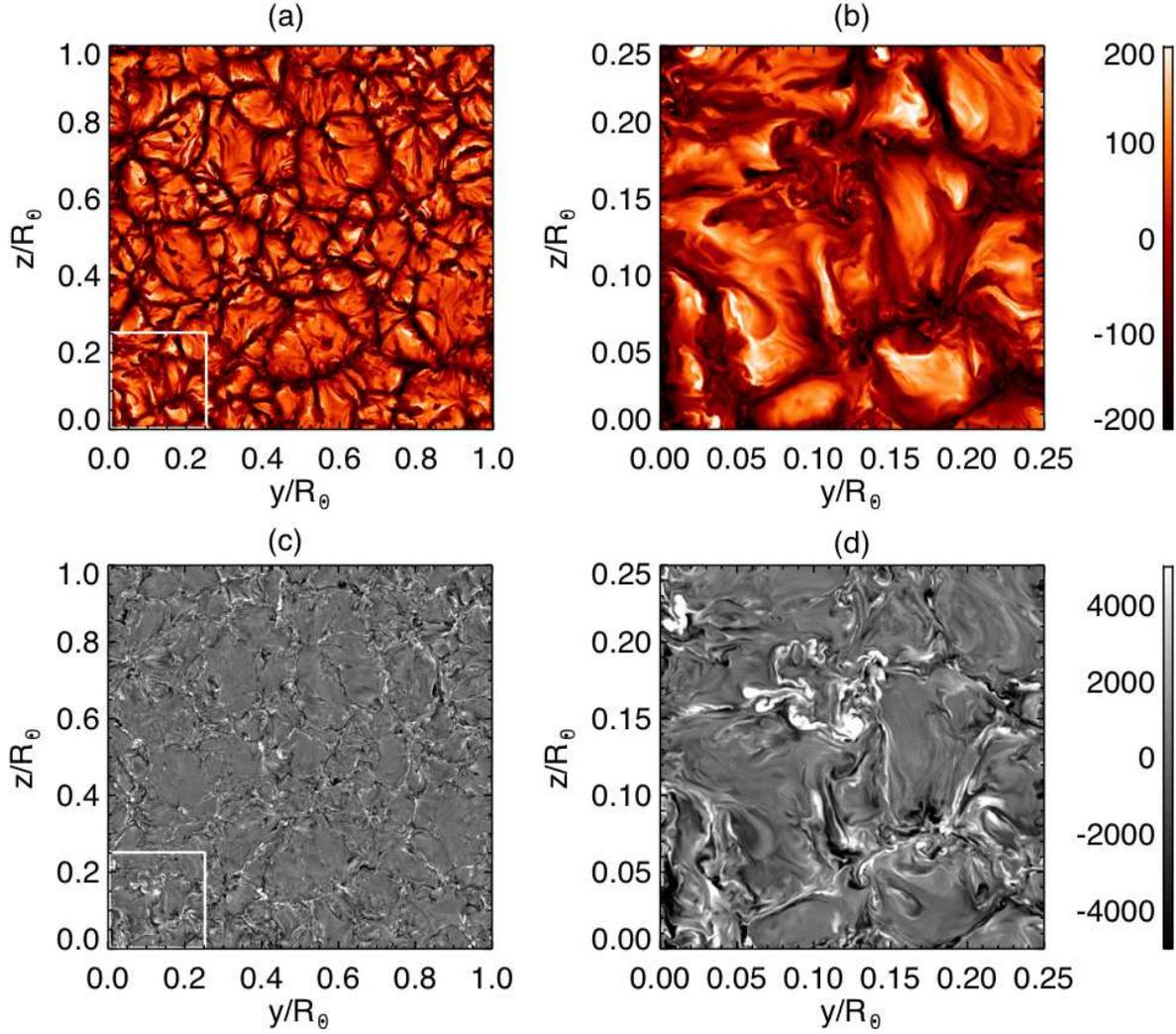}
 \caption{
 Contour of vertical velocity $v_x$ (panels a and b) in the unit of
 $\mathrm{m\ s^{-1}}$ and vertical
 magnetic field (panels c and d) in the unit of G are shown at $r=0.95R_\odot$ for case
 M2048. Panels b and d are the zoom-in figure for the panels a and c,
 respectively.
 \label{M2048vxbx95}}
\end{figure}
\begin{figure}[htbp]
 \centering
 \includegraphics[width=16cm]{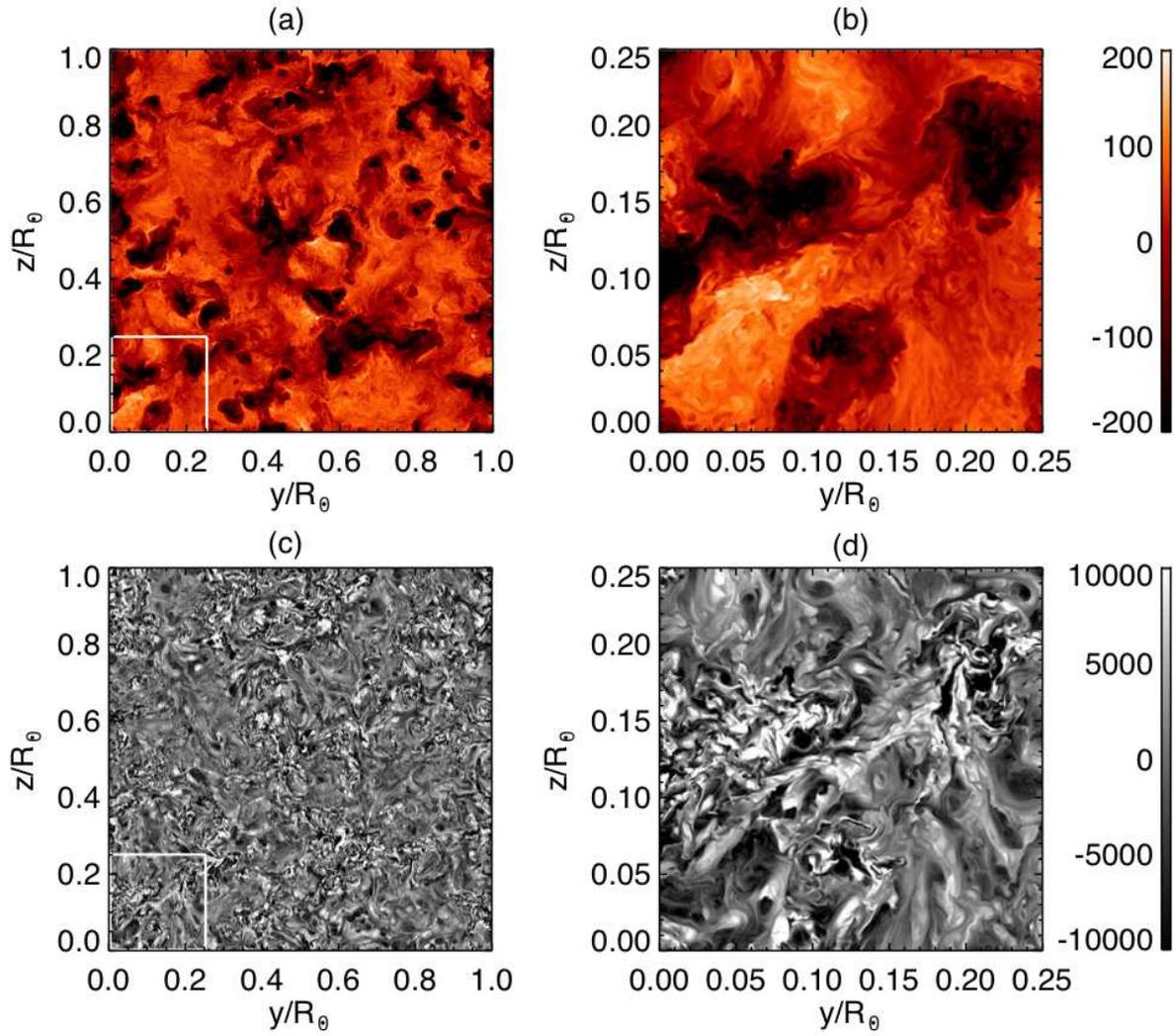}
 \caption{
 The same format as Fig. \ref{M2048vxbx95} at $r=0.8R_\odot$.
 \label{M2048vxbx80}}
\end{figure}
\begin{figure}[htbp]
 \centering
 \includegraphics[width=16cm]{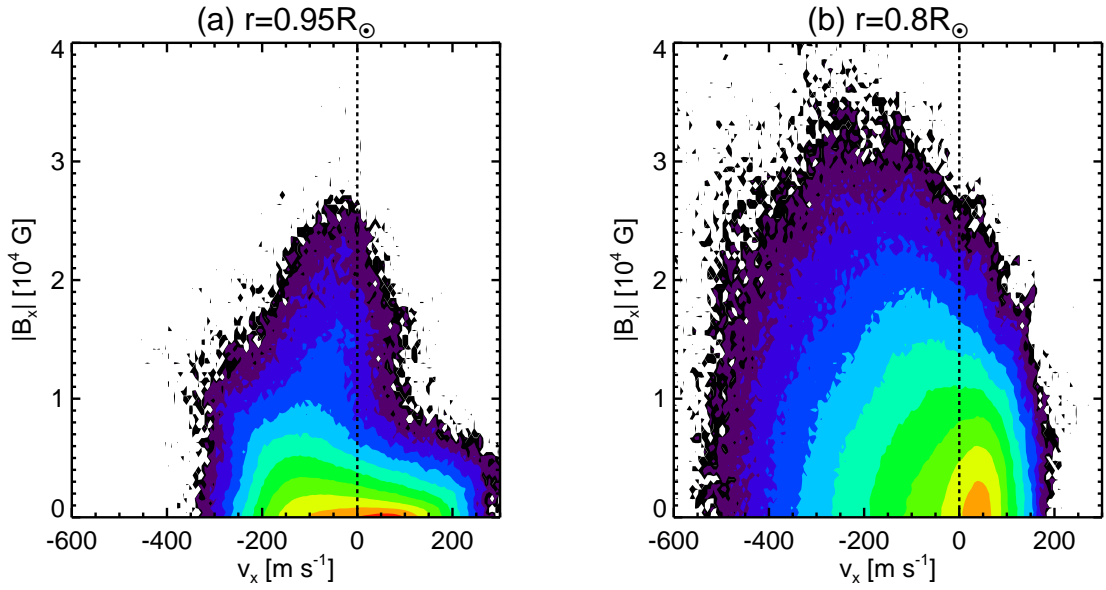}
 \caption{
 Joint PDFs of vertical velocity ($v_x$) in the unit of $\mathrm{m\
 s^{-1}}$ and vertical magnetic field
 ($B_x$) in the unit of G in case M2048 at (a) $r=0.95R_\odot$ and (b) $r=0.8R_\odot$.
 \label{M2048_2dpdfvxbx}}
\end{figure}

Although we calculate the cases M512, and M1024
for 900 days, no significant change can be seen after 300 days. We
calculate case M2048 for 300 days and compare the results at the time
of 280-300
days. Fig. \ref{comp_magene} shows the temporal evolution of magnetic
energy averaged over the whole calculation domain. As stated in
\cite{2014ApJ...789..132R}, the e-folding time scale for the growth rate
of magnetic energy
significantly depends on the resolution. The time scales are 3.24, 1.97,
and 0.99 days at case M512, M1024, and M2048, respectively. 
In addition,
the saturated magnetic energy is one order of magnitude higher than that in case
M256D. The magnetic energy in cases M1024 and M2048 converges
around $t=300\ \mathrm{days}$. 
Figs. \ref{M2048vxbx95} and
\ref{M2048vxbx80} show the distribution of vertical velocity $v_x$ and
vertical magnetic field $B_x$ at $r=0.95R_\odot$ and $r=0.8R_\odot$, respectively.
Compared with the hydrodynamic case (H2048: Fig. \ref{H2048vx}), the
appearance of the convective flow is smoother. 
The small-scale turbulence in upflow regions is strongly suppressed (see
Figs. \ref{H2048vx}d and \ref{M2048vxbx80}b).
In the upper layer, the
location of small-scale features in $v_x$ have rough coincidence with the
location of  strong vertical magnetic field $B_x$. Small-scale
magnetic field with
mixed polarity is preferentially found in downflow regions. In deeper
layers, the coincidence between downflow and strong magnetic field is
less pronounced. The magnetic field is distributed rather uniformly. 
Joint PDFs for vertical velocity ($v_x$) and vertical magnetic field
($B_x$) also show this feature in Fig. \ref{M2048_2dpdfvxbx}. While the
distribution of magnetic field smoothly connects between up- and
downflow region in the deeper layer (panel b), a
steep transition is seen near the upper boundary (panel a). 
We note
that the ratio of the RMS values of vertical magnetic field in upflow to
that in downflow ($\sim 0.5$) is not different between these layers.
\par
\begin{figure}[htbp]
 \centering
 \includegraphics[width=16cm]{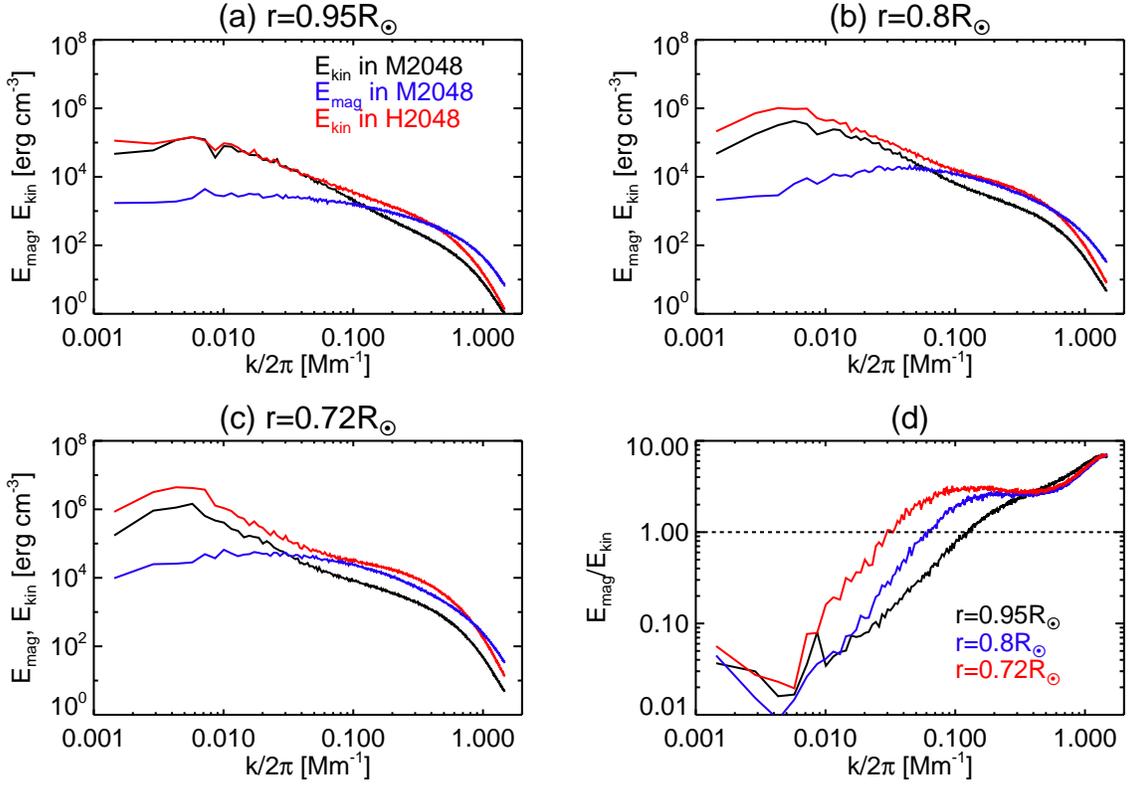}
 \caption{
Spectra in cases H2048 and M2048. Panels a, b, and c denote the result at
 $r=0.95R_\odot$, $0.8R$, and $0.72R_\odot$. The black and blue lines
 show kinetic energy, and magnetic energy in case M2048. The red line shows
 the kinetic energy in case H2048. Panel d shows the ratio of
 $E_\mathrm{mag}$ to $E_\mathrm{kin}$ at $r=0.95R_\odot$ (black),
 $0.8R_\odot$ (blue), and $0.72R_\odot$ (red).
 \label{HM2048_f}}
\end{figure}

This finding is confirmed with the spectra of
magnetic energy $E_\mathrm{mag}$ and kinetic energy $E_\mathrm{kin}$ in
Fig. \ref{HM2048_f}. The black and blue lines show $E_\mathrm{kin}$ and
$E_\mathrm{mag}$ in case M2048. The red lines show $E_\mathrm{kin}$ in
case H2048. Panels a, b, and c show the results at $r=0.95R_\odot$,
$0.8R_\odot$, and $0.72R_\odot$, respectively. Fig. \ref{HM2048_f}d
shows the ratio of $E_\mathrm{mag}$ to $E_\mathrm{kin}$. The black,
blue, and red lines show the result at $r=0.95R_\odot$, $0.8R_\odot$,
and $0.72R_\odot$, respectively. 
In every layer, super-equipartition magnetic field is found at
small scales. There is depth dependence of the scale where the magnetic
energy exceeds the kinetic energy.
The scales are 8.1, 15, and 31 Mm
for $r=0.95R_\odot$, $0.8R_\odot$, and $0.72R_\odot$, respectively. In
Paper I, it is argued that there are two factors that influence the dependence of
dynamo efficiency on the depth. One is the negative Poynting flux in the
convection zone. In the upper convection zone, magnetic energy is
rapidly transport downward and accumulates in the lower convection zone.
This makes small-scale dynamo action more challenging near the top and
less challenging near the bottom of the domain.
The other factor is the dependence of the intrinsic spatial scale of the
convection flow on the depth, i.e., the convection in deeper layer has
larger spatial scale. For a fixed grid spacing convective structures are
better resolved near the bottom of the domain.
The super-equipartition magnetic energy is 6 times larger than kinetic
energy in the diffusion scale ($k/2\pi>0.5\ \mathrm{Mm^{-1}}$). This likely
indicates a numerical magnetic Prandtl number larger than unity in our
calculation.
In upper layer
($r=0.95R_\odot$: Fig. \ref{HM2048_f}a), small-scale kinetic
energy is selectively suppressed by magnetic field and the larger scale kinetic
energy remains. On the other hand, all the scales are suppressed in the
deeper layer at $r=0.72R_\odot$ (Fig. \ref{HM2048_f}c).\par
\begin{figure}[htbp]
 \centering
 \includegraphics[width=16cm]{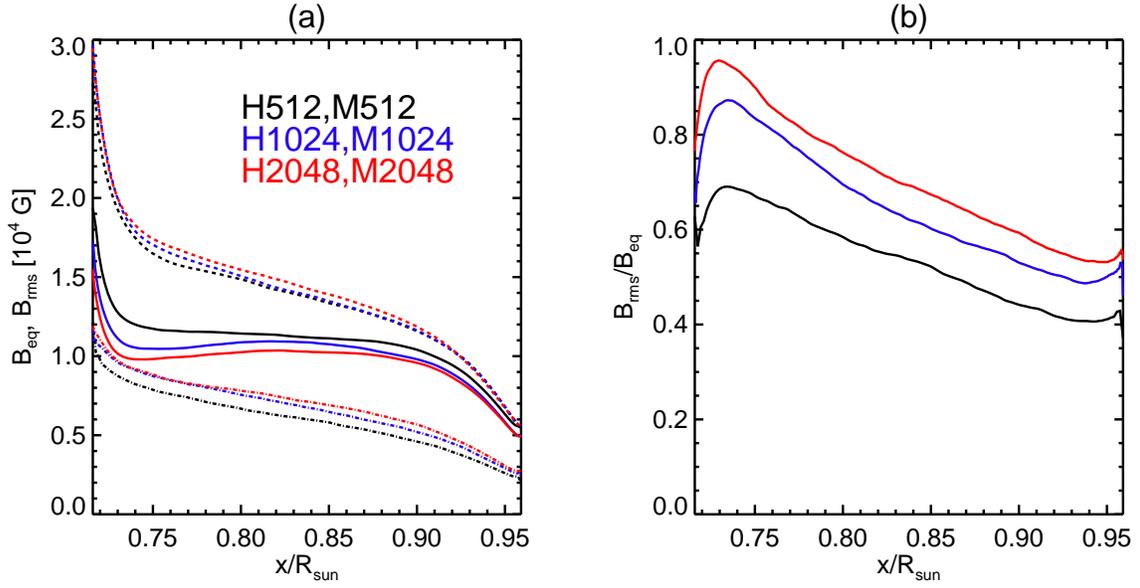}
 \caption{The relation between equipartition magnetic field
 $B_\mathrm{eq}$ and RMS magnetic field $B_\mathrm{rms}$. 
(a) The solid and dash-dotted lines show $B_\mathrm{eq}$ and
 $B_\mathrm{rms}$, respectively for the cases M512 (black), M1024
 (blue), and M2048 (red). The dotted black, blue, and red lines show
 $B_\mathrm{eq}$ in cases H512, H1024, and H2048, respectively.
(b) The ratio of $B_\mathrm{rms}$ to $B_\mathrm{eq}$ in cases M512,
 M1024, and M2048.
 \label{comp_beq}}
\end{figure}

Fig. \ref{comp_beq}a shows $B_\mathrm{eq}$ and $B_\mathrm{rms}$. 
The solid and dash-dotted lines show $B_\mathrm{eq}$ and
$B_\mathrm{rms}$ in case M512 (black), M1024 (blue), and M2048 (red),
respectively. The dotted lines show $B_\mathrm{eq}$ in hydrodynamic
cases H512 (black), H1024 (blue), and H2048 (red). Although around the
base of
the convection zone, the strength of the magnetic field between cases
M1024 and M2048 converges, the reduction of $B_\mathrm{eq}$ is larger in
M2048. The higher resolution simulation has more kinetic energy  on small scales
(recall the depression in Fig. \ref{comp_f_hydro}c). Since the Lorentz
feedback has a larger effect at smaller scales, the reduction of
$B_\mathrm{eq}$ is larger in case M2048.
Fig. \ref{comp_beq}b
shows the ratio of $B_\mathrm{rms}$ to $B_\mathrm{eq}$ in cases M512,
M1024, and M2048. At the base of the convection zone, case M2048
achieves 95\% of the equipartition field strength. \par
\begin{figure}[htbp]
 \centering
 \includegraphics[width=16cm]{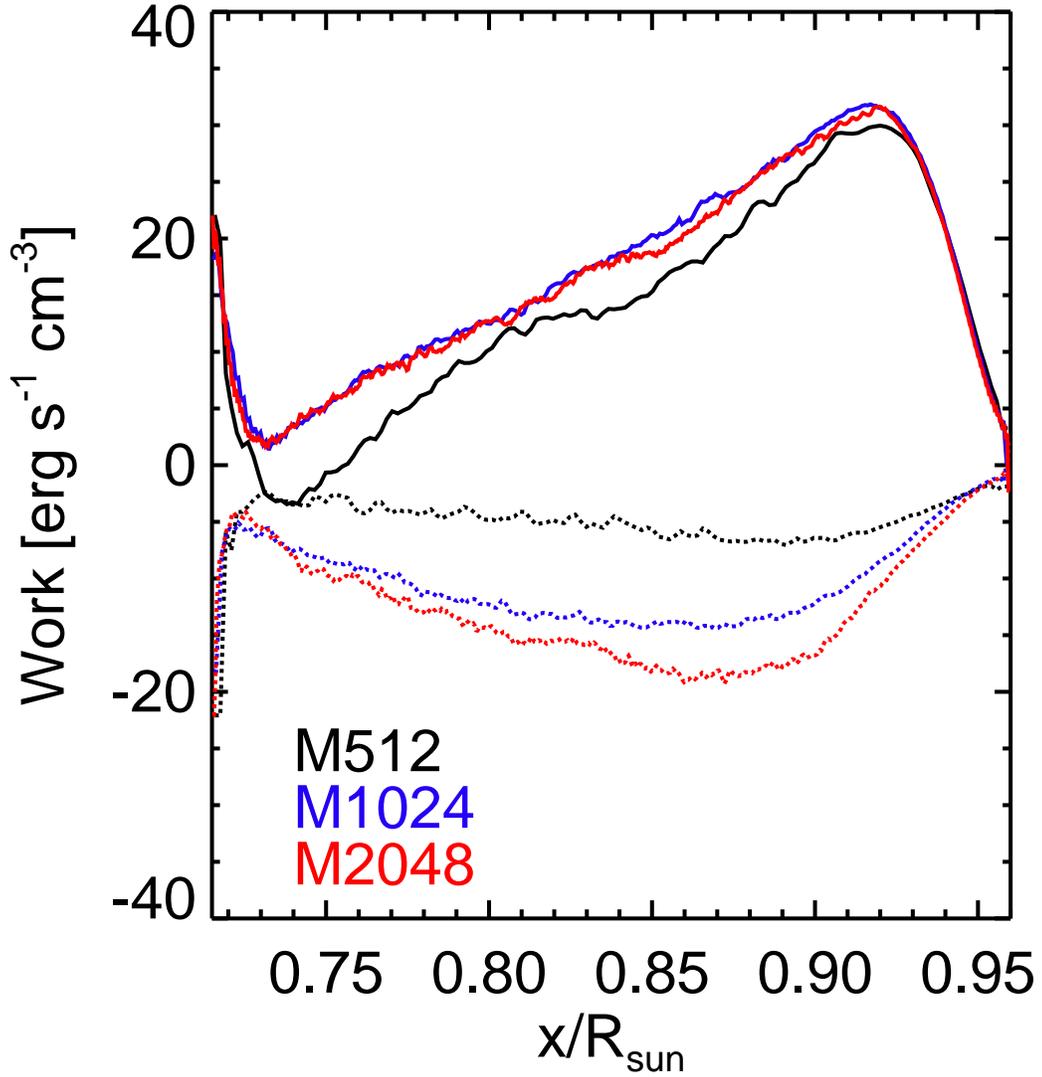}
 \caption{ The work densities of pressure/buoyancy ($W_\mathrm{d}$:
 solid line) and
 the Lorentz force ($W_\mathrm{l}$: dotted line) (see eqs. (\ref{pressure}) and
 (\ref{lorentz})). The black, blue and red lines show the results of
 M512, M1024, and M2048, respectively.
 \label{work}}
\end{figure}

Fig. \ref{work} shows the horizontally averaged work densities of
pressure/buoyancy ($W_\mathrm{d}$) and the Lorentz force ($W_\mathrm{l}$)
defined as 
\begin{eqnarray}
 W_\mathrm{d} &=&
  -\frac{\int {\bf v}\cdot(\nabla p_1 + \rho_1 g{\bf e_x})dS}{\int dS},
  \label{lorentz}\\
 W_\mathrm{l} &=&
  \frac{\int {\bf v}\cdot({\bf j}\times{\bf B})dS}{\int dS},
  \label{pressure}
\end{eqnarray}
where $S$ is the horizontal plane. The work by pressure and buoyancy 
almost converges between M1024 and M2048. The conversion rate
from kinetic energy to magnetic energy ($W_\mathrm{l}$) is 36\%, 61\%,
and 76\% of the pressure/buoyancy work ($W_\mathrm{d}$) in the
calculation domain in cases M512, M1024, and M2048, respectively.
The magnetic Prandtl number is close to unity in this
study since we only use numerical diffusivity. Recently 
\citet{2011AN....332...51B,2014ApJ...791...12B}
found that for efficient dynamos
with $\mathrm{Rm}\gg 1$ the work by the Lorentz force and consequently the ratio of
viscous to resistive energy dissipation is dependent on the magnetic Prandtl
number. In particular in the regime $1 \ll \mathrm{Rm} \ll \mathrm{Re}$, which is most relevant 
for solar and stellar convection zones most of the energy dissipation happens 
through resistivity. This means that the value $W_\mathrm{l}/W_\mathrm{d}$ increases 
with lowering the magnetic Prandtl number and should be close to unity (i.e. the
small-scale dynamo is maximally efficient). Note that in global dynamo simulations with
moderate $\mathrm{Rm}$, where an efficient small-scale dynamo is not present, the value
$W_\mathrm{l}/W_\mathrm{d}$ (mostly due to a large-scale dynamo) can depend on
other parameters such as Rossby number \citep{2013MNRAS.431L..78S}.
\par
\begin{figure}[htbp]
 \centering
 \includegraphics[width=16cm]{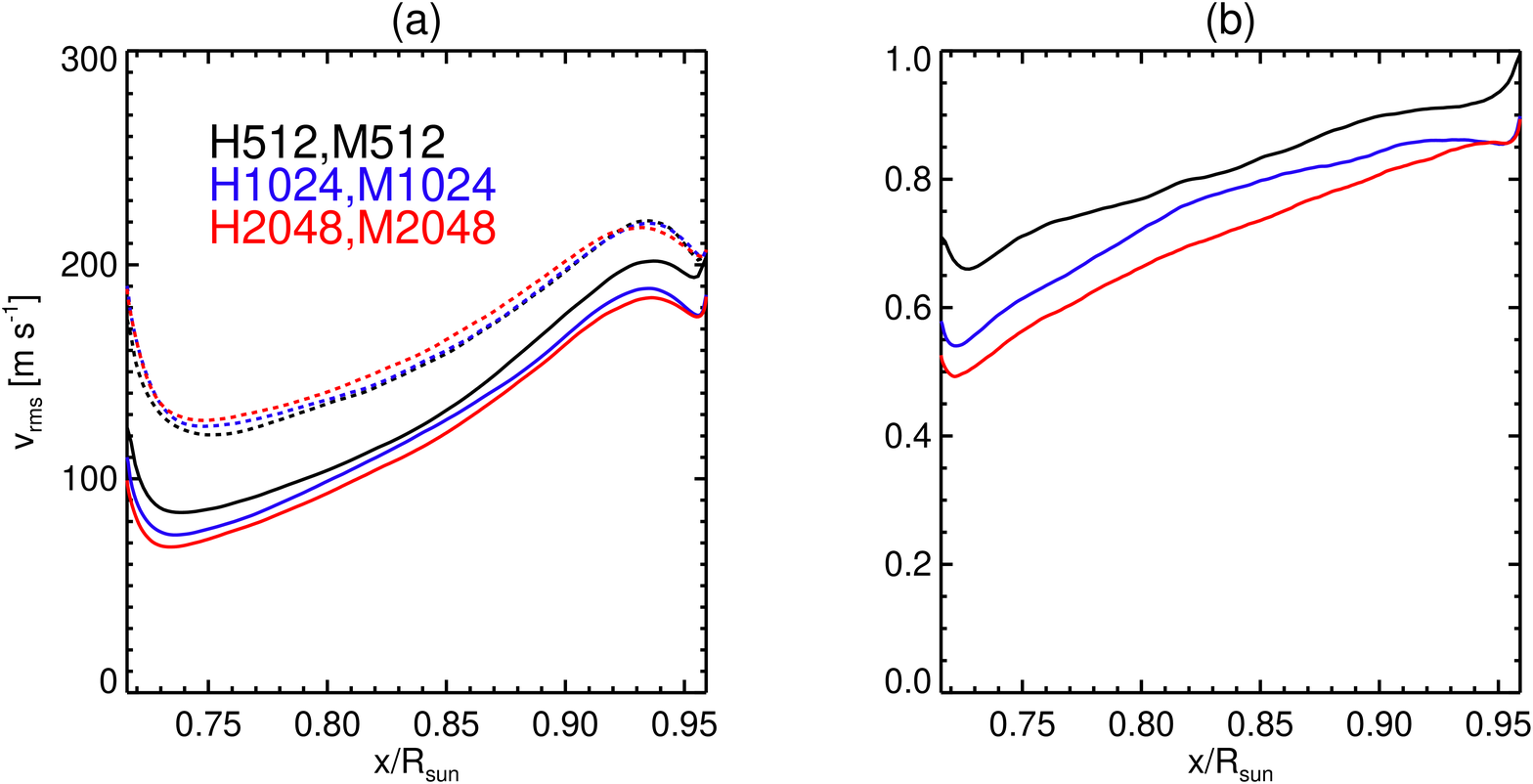}
 \includegraphics[width=16cm]{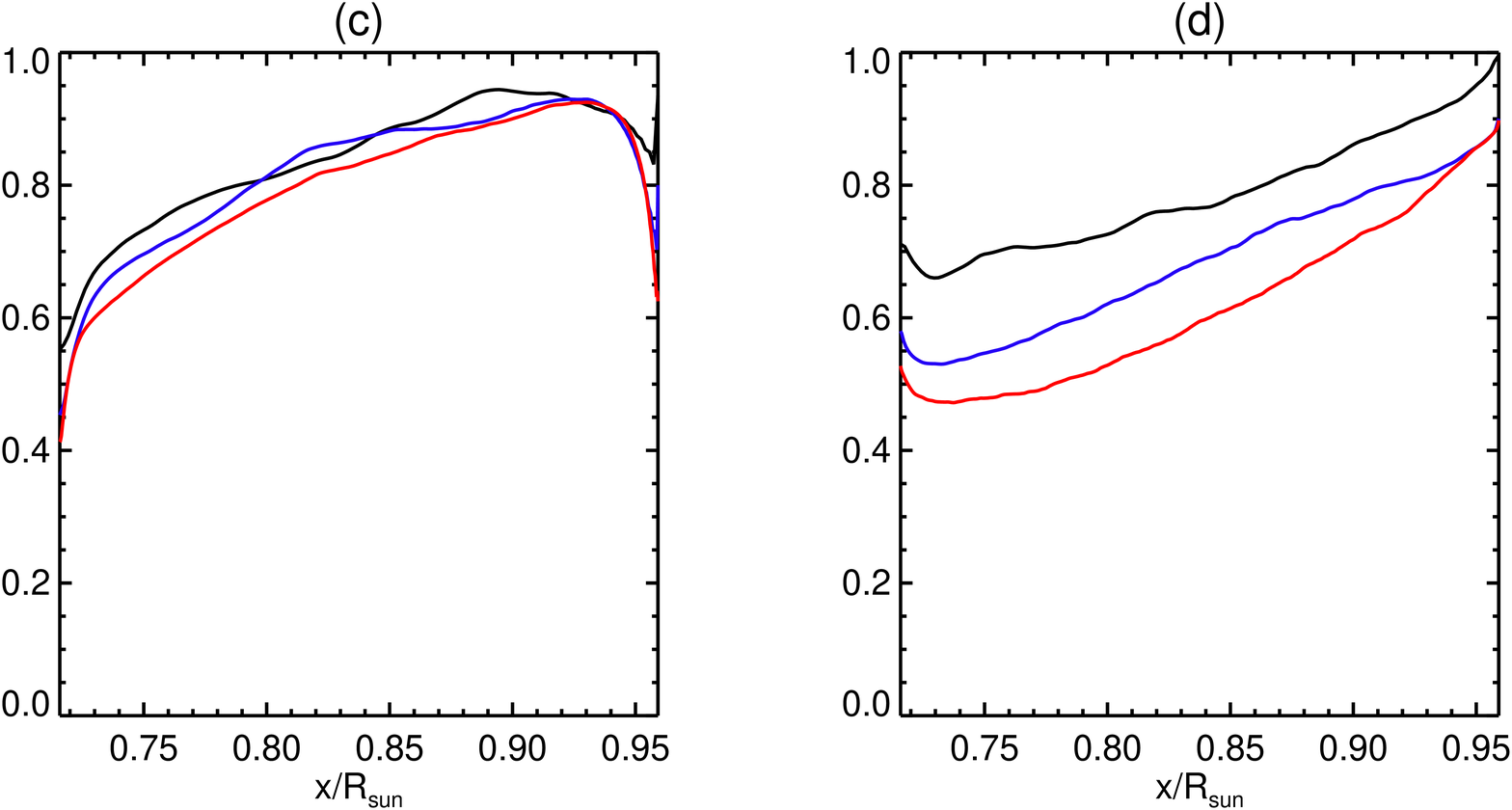}
 \caption{
The solid (dotted) black, blue and red lines show the results in cases
 M512(H512), M1024(H1024), and M2048(H2048), respectively. (a) The
 distribution of RMS velocities. 
(b) The ratio of total RMS velocity $v_\mathrm{RMS}$ in the MHD cases to hydrodynamic cases,
(c) The ratio of vertical RMS velocity $v_{x\mathrm{RMS}}$ in the MHD cases to
 hydrodynamic cases, and
(d) The ratio of horizontal RMS velocity $v_\mathrm{h\mathrm{RMS}}$
 ($v_h=\sqrt{v_y^2+v_z^2}$) in MHD cases to hydrodynamic cases.
 \label{comp_rmsvel}}
\end{figure}

\begin{figure}[htbp]
 \centering
 \includegraphics[width=16cm]{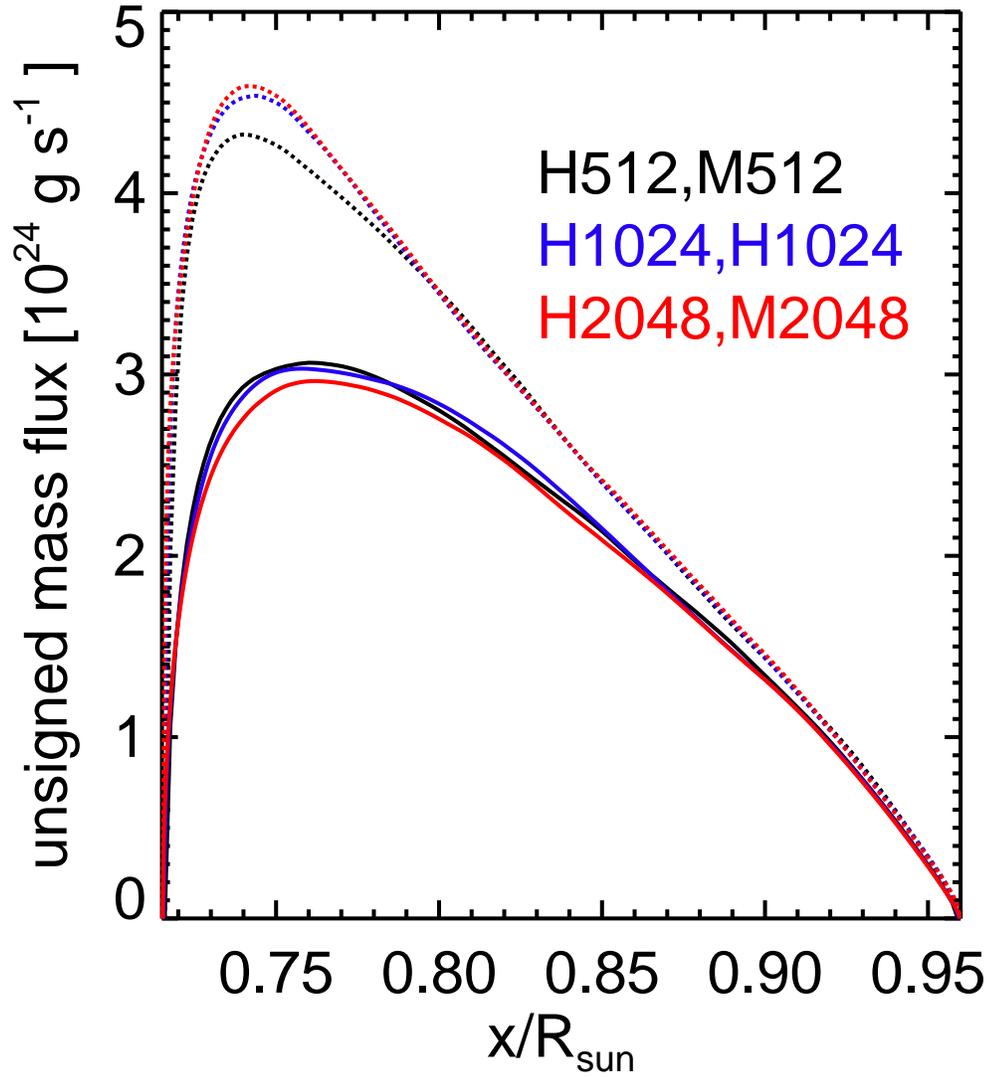}
 \caption{Horizontally integrated unsigned mass flux $\rho_0 |v_x|$.
 The solid (dotted) black, blue and red lines show the results in cases
 M512(H512), M1024(H1024), and M2048(H2048), respectively.
 \label{unsign_massflux}}
\end{figure}

 
Fig. \ref{comp_rmsvel} shows the RMS velocity for all cases. In panel a, the
solid (dotted), black, blue, and red lines show the results in cases M512
(H512), M1024 (H1024), and M2048 (H2048). Figs. \ref{comp_rmsvel}b, c,
and d show the ratio of RMS velocity in the MHD case to that in
the hydrodynamic case. The black, blue and red lines show the result in
cases H(M)512, H(M)1024 and H(M)2048. The panels b, c, and d show the
RMS velocities including all three components, the x-component, and
the horizontal component ($v_\mathrm{h}=\sqrt{v_x^2+v_y^2}$),
respectively. As shown in Fig. \ref{comp_beq},
Fig. \ref{comp_rmsvel} also shows that the reduction of the velocity
with magnetic field does not converge with M1024 and M2048. In the
highest resolution case M2048, the RMS velocity is reduced to 85\% and
50\% at the top and bottom boundaries. Figs. \ref{comp_rmsvel}c and d
show that horizontal velocity is more suppressed by magnetic field, which is
a key feature for understanding the energy transport in the following
discussion.
While the suppression of $v_x$ is converged, the suppression of $v_\mathrm{h}$ shows strong
resolution dependence.
As shown in the spectra (Fig. \ref{HM2048_f}b and c), the kinetic energy
is reduced independently of
the spatial scale from the middle to the base of the convection zone. As
a result, the unsigned mass flux is also suppressed in the deeper region
(Fig. \ref{unsign_massflux}). This indicates that the amount of over turning
is reduced. It is thought that this is caused by the large-scale
magnetic field component, since the reduction does not depend on the
resolution significantly.
\par
\begin{figure}[htbp]
 \centering
 \includegraphics[width=16cm]{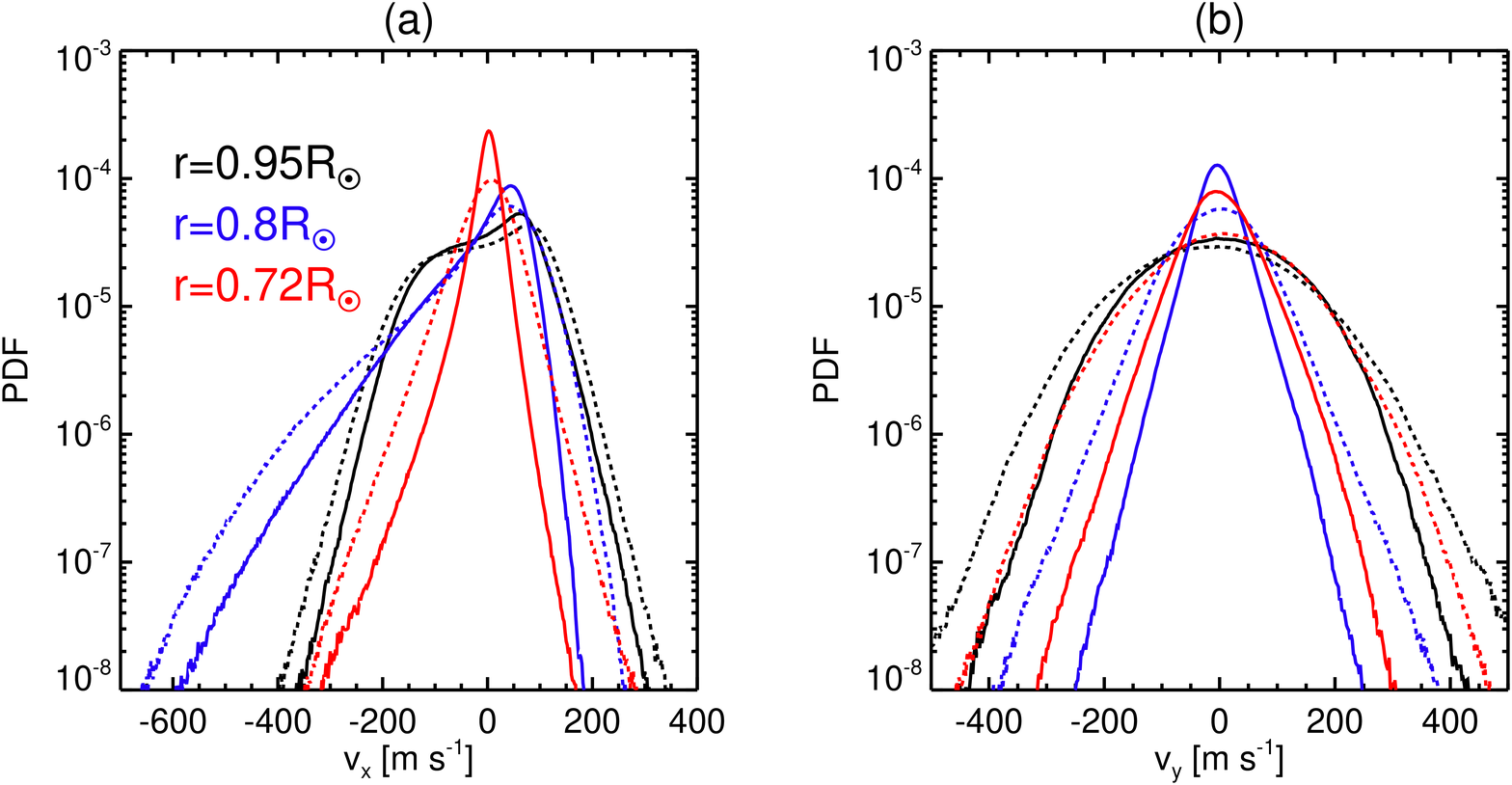}
 \includegraphics[width=16cm]{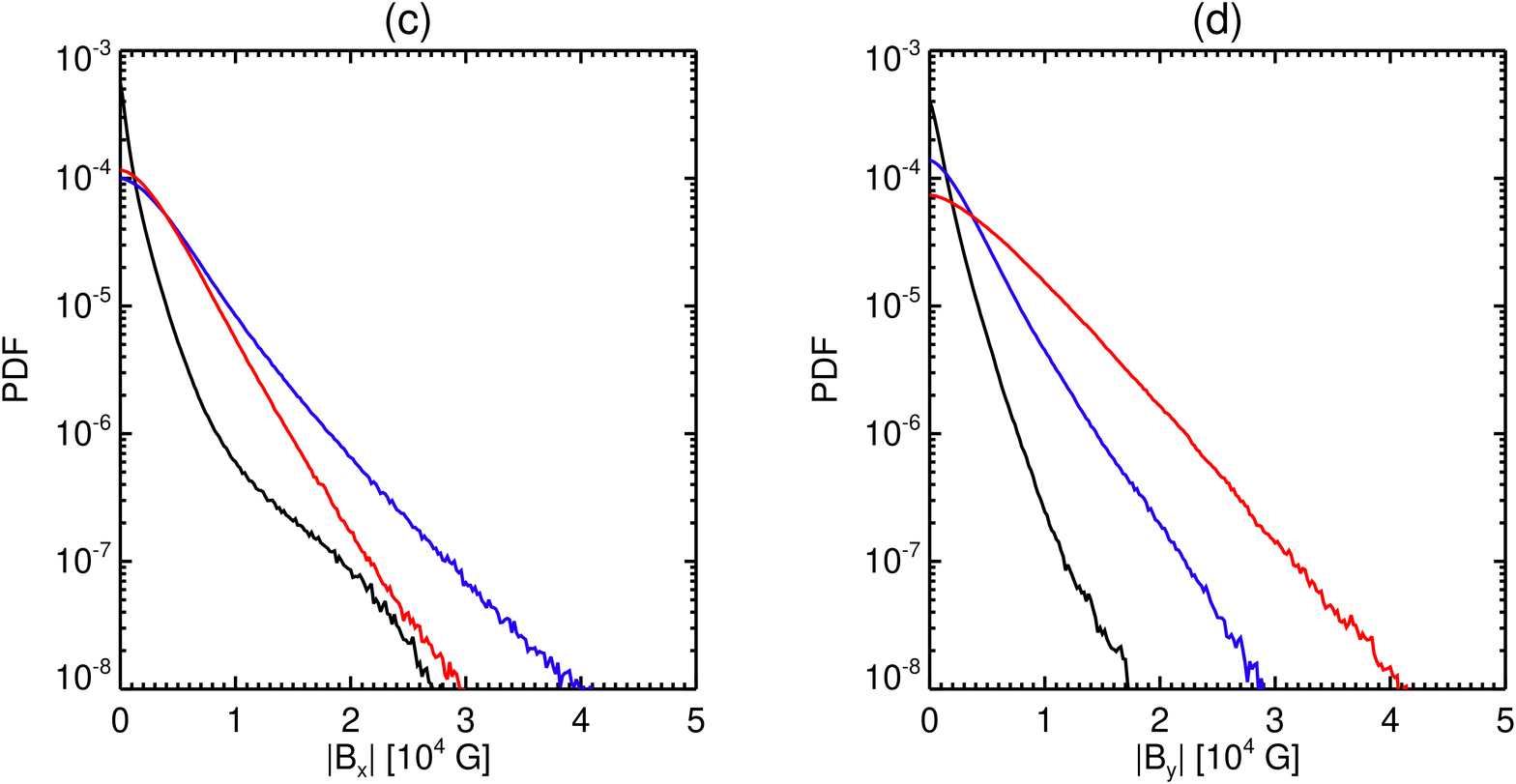}
 \caption{
PDFs for (a) the vertical velocity $v_x$, (b) a horizontal velocity
 $v_y$, (c) the vertical magnetic field $B_x$ and (d) a horizontal
 magnetic field $B_y$. The black, blue and red lines show the result at
  $r=0.95R_\odot$, $0.8R_\odot$, and $0.72R_\odot$, respectively.
The dotted and solid show the result in cases H2048, and M2048.
 \label{HM2048_pdf}}
\end{figure}
\begin{figure}[htbp]
 \centering
 \includegraphics[width=16cm]{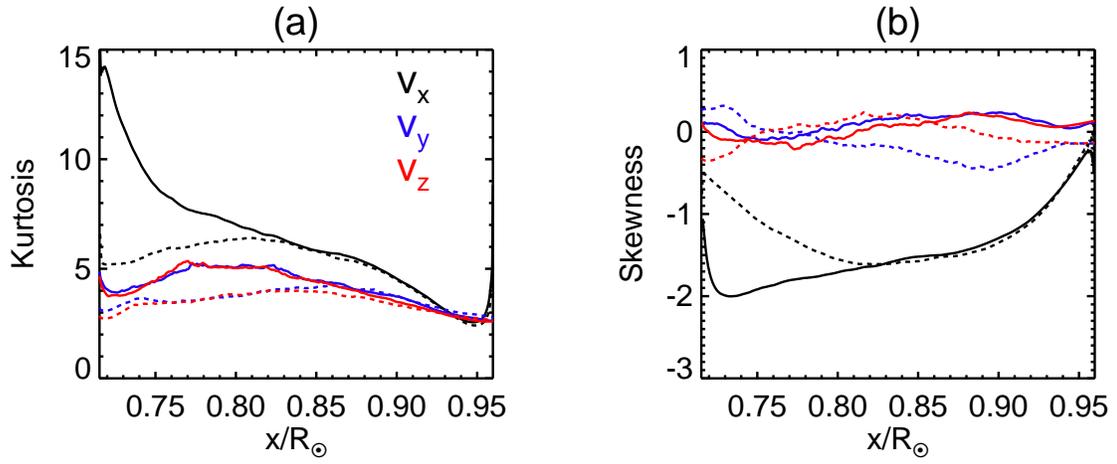}
 \caption{
 (a) Kurtosis and (b) skewness of the velocities. Black, blue and red
 lines show $v_x$, $v_y$, and $v_z$, respectively. Dotted and solid
 lines show the results of H2048 and M2048, respectively.
 \label{ks_pdf1}}
\end{figure}

Figs. \ref{HM2048_pdf}a, and b show PDFs for $v_x$, and $v_y$,
respectively.
The dotted and solid lines show the results in cases
H2048 and M2048. The black, blue and red lines show the result at
$r=0.95R_\odot$, $0.8R_\odot$, and $0.72R_\odot$, respectively. As shown
in Paper I, the vertical velocity has a asymmetric distribution and the
distribution of the horizontal velocity is almost  Gaussian.
Larger velocity values are more strongly reduced at all depths.
Fig. \ref{ks_pdf1} shows the kurtosis $\mathcal{K}$ and skewness
$\mathcal{S}$, which show intermittency and asymmetry of distribution,
respectively.
As explained above, the horizontal velocity has almost Gaussian
distribution ($\mathcal{K}\sim3$ and $\mathcal{S}\sim0$) and vertical
velocity ($v_x$) has higher intermittency and asymmetry in the
hydrodynamic case (H2048: dotted lines). When the magnetic field is
included (M2048: solid line), intermittency is increased especially at
the bottom part of the convection zone. This indicates that the higher
velocity is selectively suppressed by the magnetic field.
Figs. \ref{HM2048_pdf}c and d show PDFs for $B_x$, and $B_y$,
respectively. The PDFs of magnetic field are similar to those in
\cite{2014ApJ...789..132R}. Due to the boundary condition at the bottom,
the vertical magnetic field at $r=0.8R_\odot$ is larger than
that in $r=0.72R_\odot$. The peak magnetic field strength is around 65
kG near the base of the convection zone. Although we do not generate flux
concentrations that are large enough to lead to flux emergence, the achieved magnetic field is
sufficiently strong in order to reproduce the active region features
\citep{2011ApJ...741...11W,2012ApJ...759L..24H}. The vertical field
around the top boundary (black line in Fig. \ref{HM2048_pdf}c) shows an
extended tail. Although this is also seen in photospheric simulations,
our result 
is likely strongly influenced
by the impenetrable boundary
condition at the top.\par
\subsection{Energy transport in high resolution calculations}
\begin{figure}[htbp]
 \centering
 \includegraphics[width=16cm]{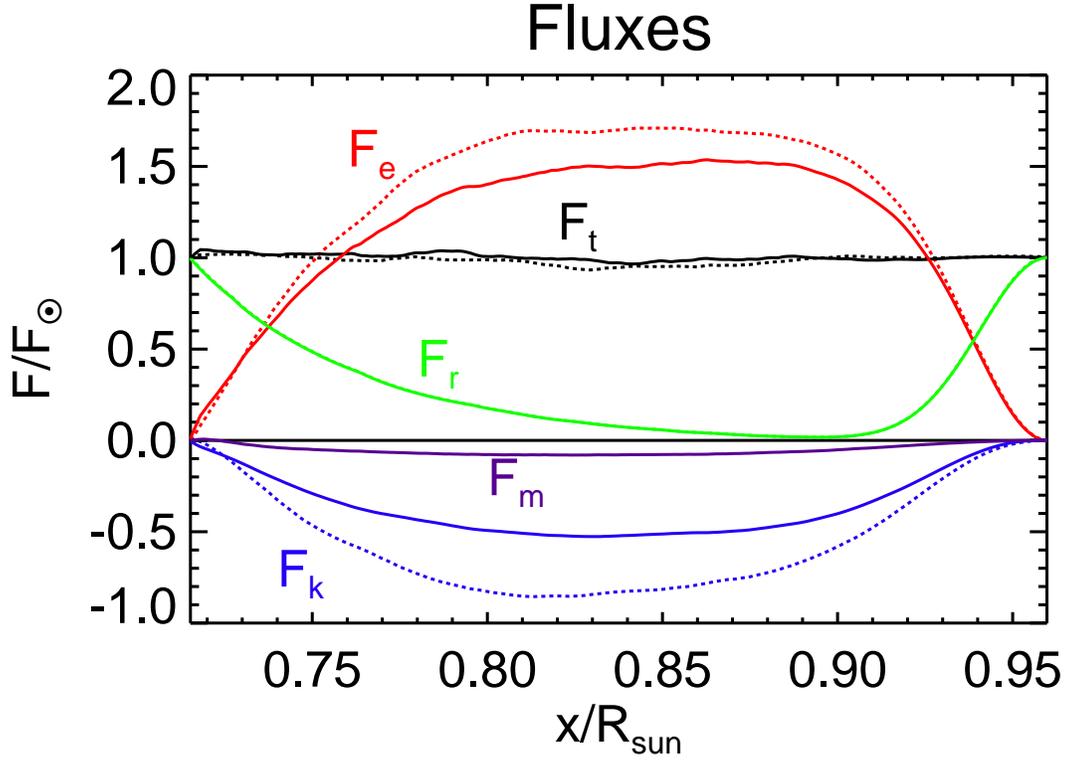}
 \caption{
Energy fluxes in cases H2048 and M2048 (dotted and solid
 lines, respectively). The black, blue, red, light green, and purple
 lines show the total, kinetic, enthalpy, radiative, and Poynting
 fluxes, respectively. Black solid line shows the $F=0$ axis for a reference.
The solar flux in this study is $F_\odot=1.21\times10^{11}\ \mathrm{erg\ 
 s^{-1}\ cm^{-2}}$.
 \label{HM2048_flux}}
\end{figure}

\begin{figure}[htbp]
 \centering
 \includegraphics[width=16cm]{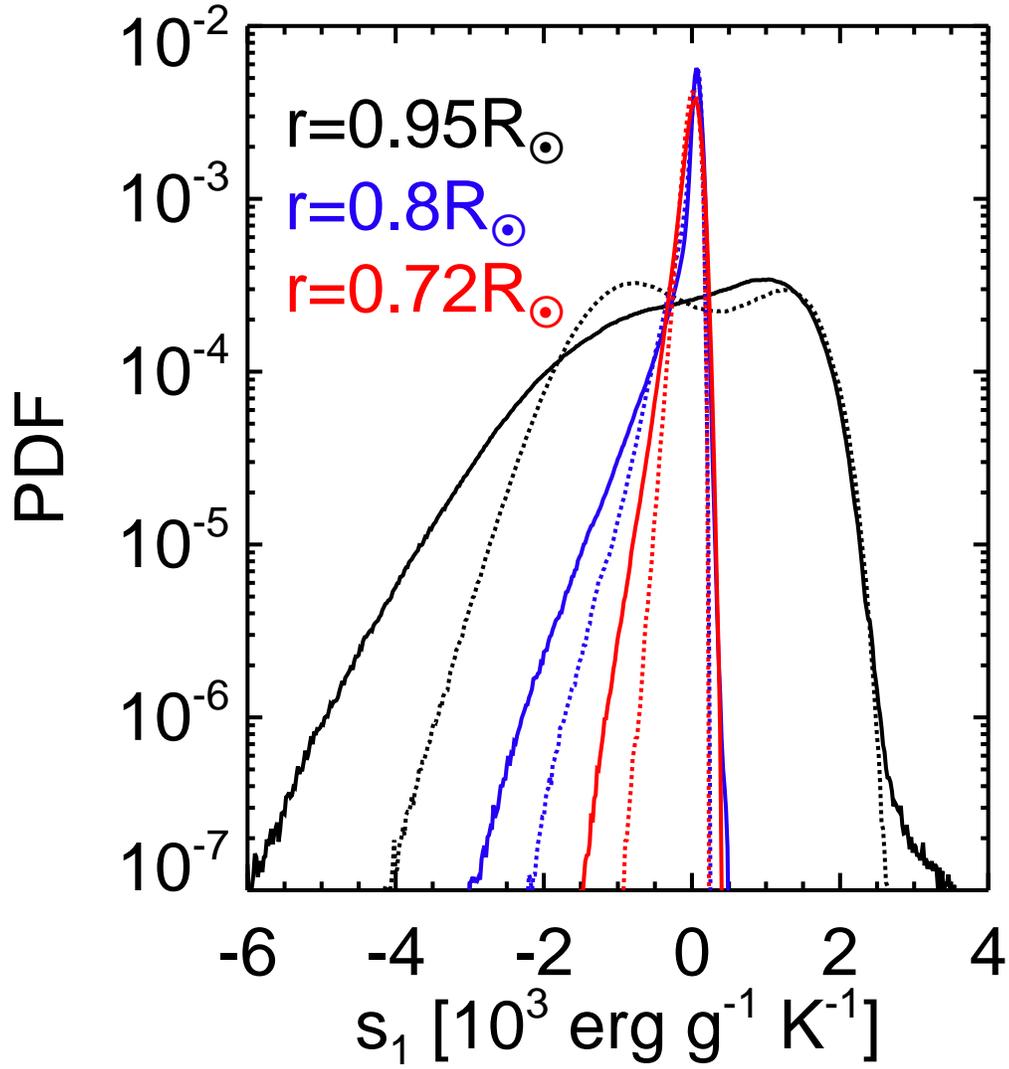}
 \caption{
PDFs of the entropy. The black, blue and red lines show the result at
 $r=0.95R_\odot$, $0.8R_\odot$, and $0.72R_\odot$. The dotted and solid
 lines show the results in cases H2048 and M2048, respectively.
 \label{HM2048_pdf_ent}}
\end{figure}

\begin{figure}[htbp]
 \centering
 \includegraphics[width=16cm]{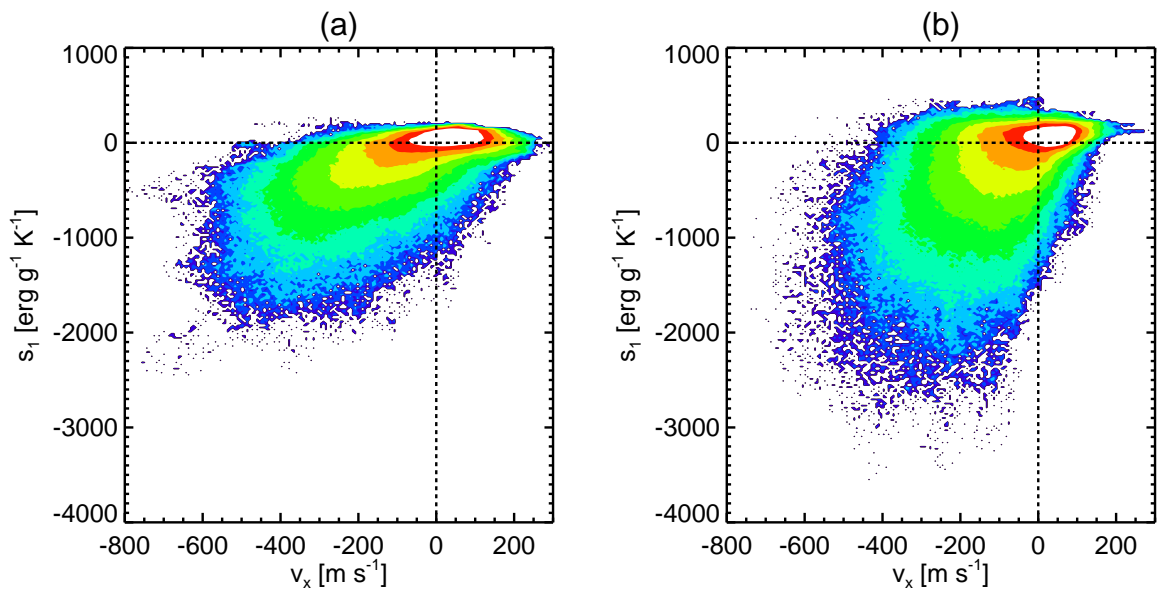}
 \caption{
 Joint PDFs with the vertical velocity $v_x$ and the entropy $s_1$ at
 $r=0.80R_\odot$. The panels a and b shows the result in cases H2048 and
 M2048, respectively.
 \label{HM2048_2dpdf}}
\end{figure}

\begin{figure}
 \centering
 \includegraphics[width=16cm]{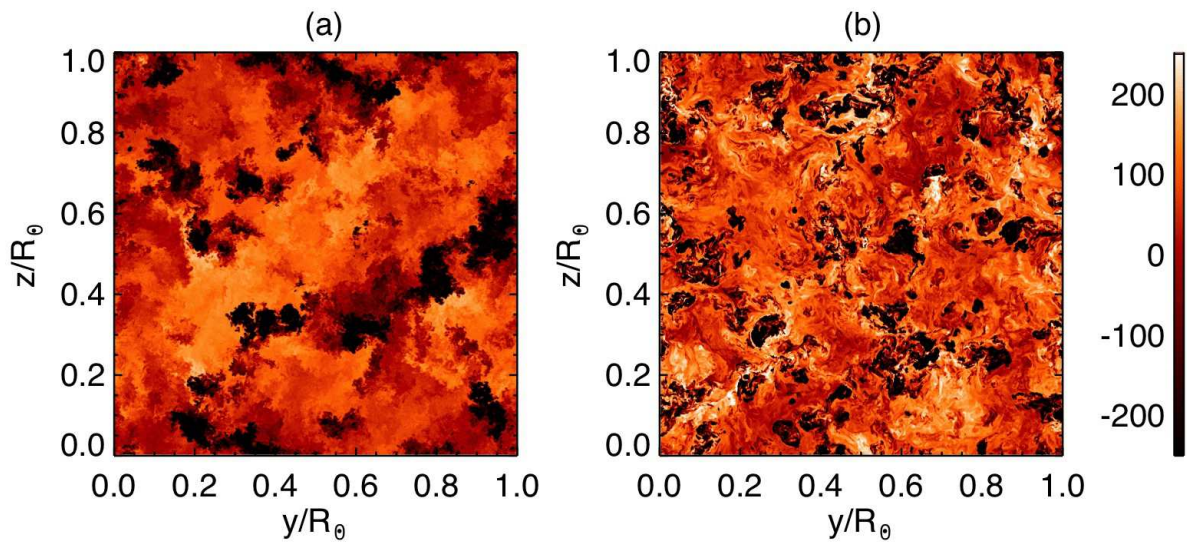}
 \caption{
 Contour of entropy at $r=0.8R_\odot$ in cases (a) H2048 and (b) M2048
in the unit of $\mathrm{erg\ g^{-1}\ K^{-1}}$
 \label{HM2048se}
 }
\end{figure}

\begin{figure}
 \centering
 \includegraphics[width=16cm]{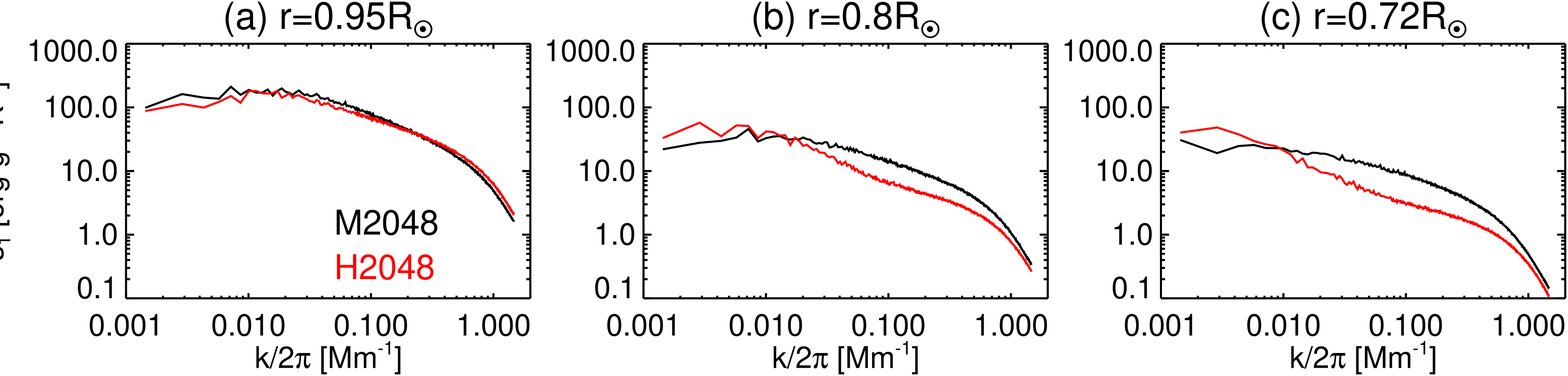}
 \caption{
 Spectra of entropy at (a) $r=0.95R_\odot$, (b) $r=0.8R_\odot$, and (c)
 $r=0.72R_\odot$. The red and black lines show the result in cases
 H2048 and M2048, respectively.
 \label{HM2048_f_entropy}
 }
\end{figure}

\begin{figure}
 \centering
 \includegraphics[width=16cm]{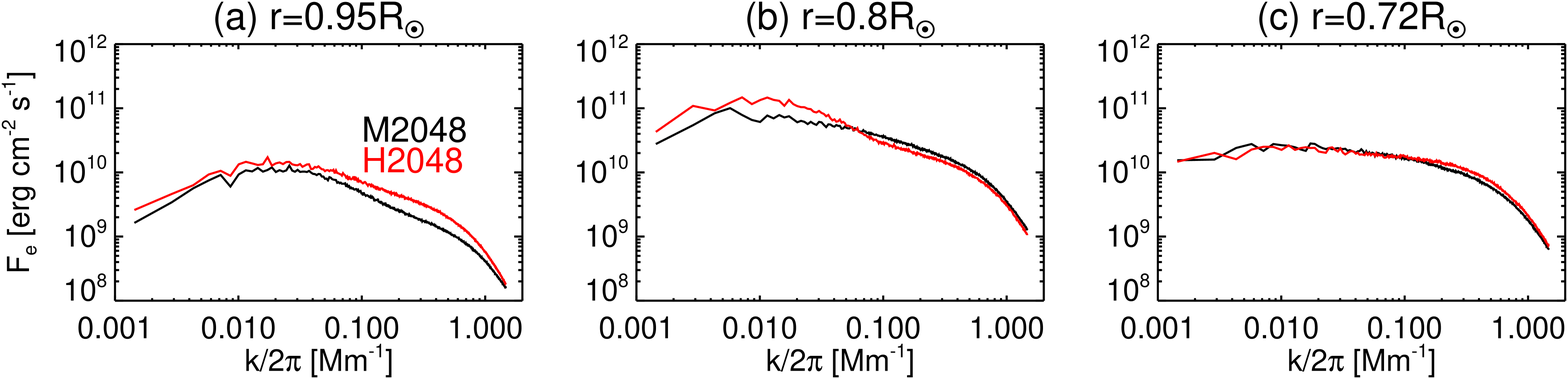}
 \caption{
 Spectra of enthalpy flux at (a) $r=0.95R_\odot$, (b) $r=0.8R_\odot$, and (c)
 $r=0.72R_\odot$. The red and black lines show the result in cases
 H2048 and M2048, respectively.
 \label{HM2048_f_enthalpy}
 }
\end{figure}

\begin{figure}
 \centering
 \includegraphics[width=16cm]{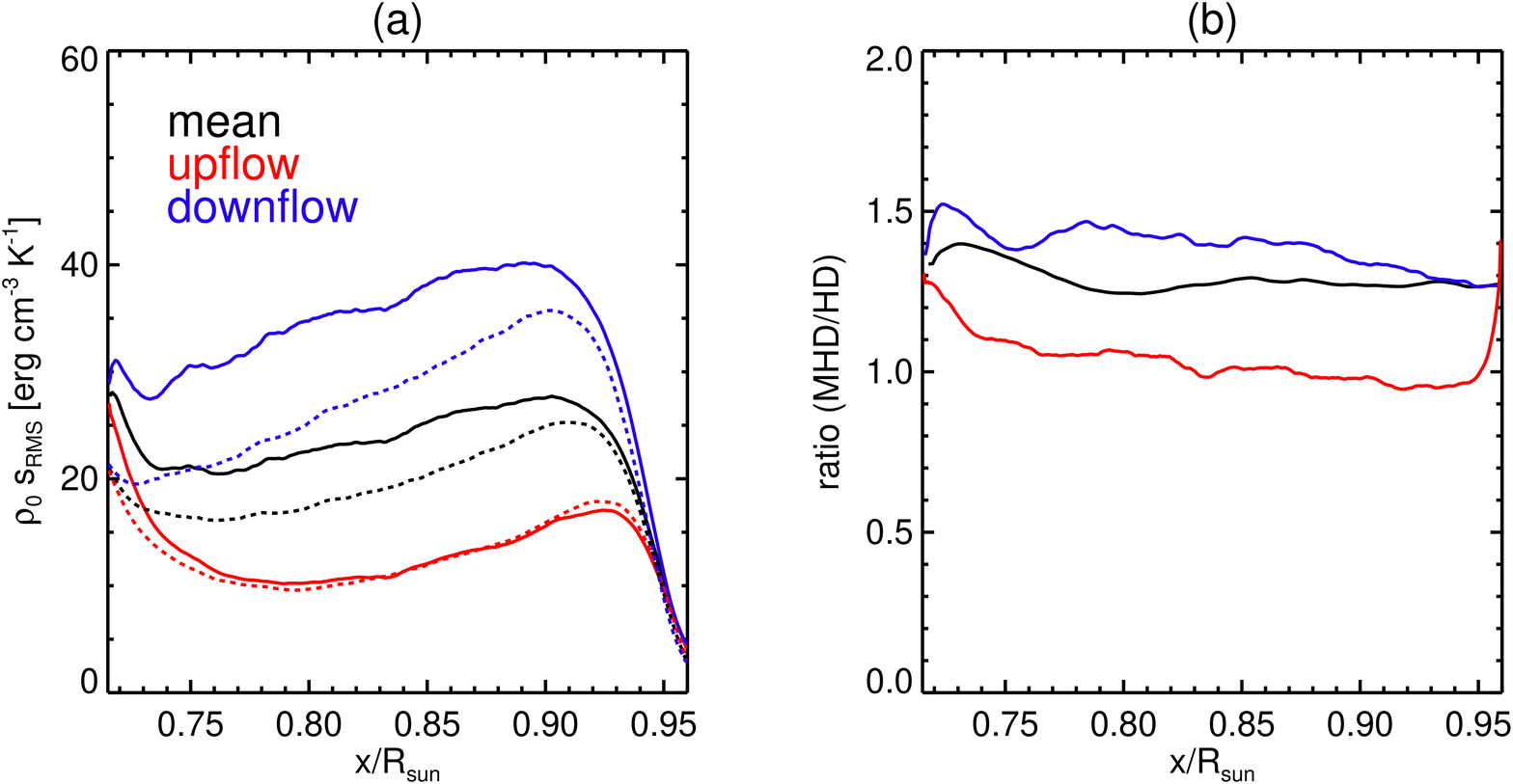}
 \caption{
(a) The RMS values of entropy perturbation. Black, red, and blue lines show the mean
 value, value in upflow and downflow, respectively. Dotted and solid
 lines show the results in cases H2048 and M2048, respectively.
(b) The ratio of the RMS entropy perturbation between the cases H2048 and M2048.
 \label{entropy_updown}
 }
\end{figure}

\begin{figure}
 \centering
 \includegraphics[width=16cm]{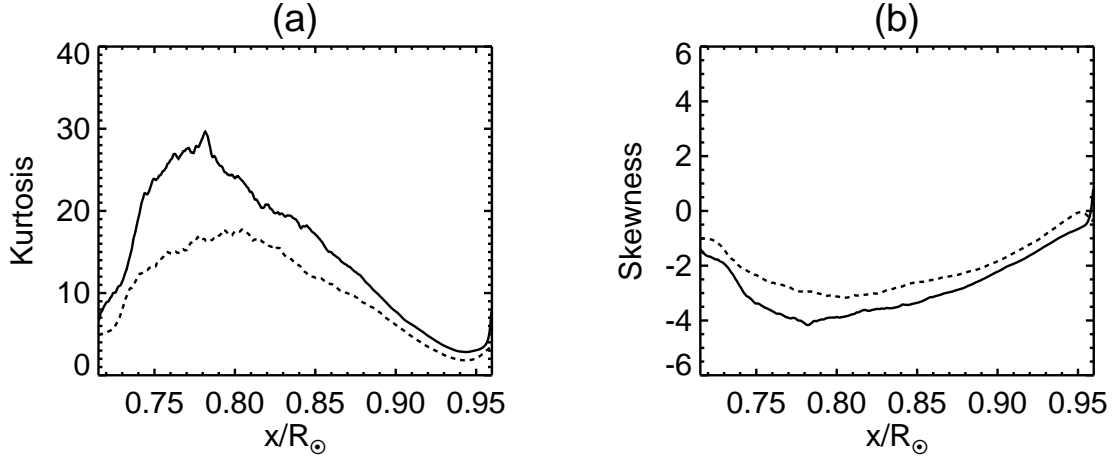}
 \caption{
 (a) Kurtosis and (b) Skewness of the entropy. Dotted and solid lines
 show the result in cases H2048 and M2048, respectively.
 \label{ks_pdf2}
 }
\end{figure}

\begin{figure}
 \centering
 \includegraphics[width=16cm]{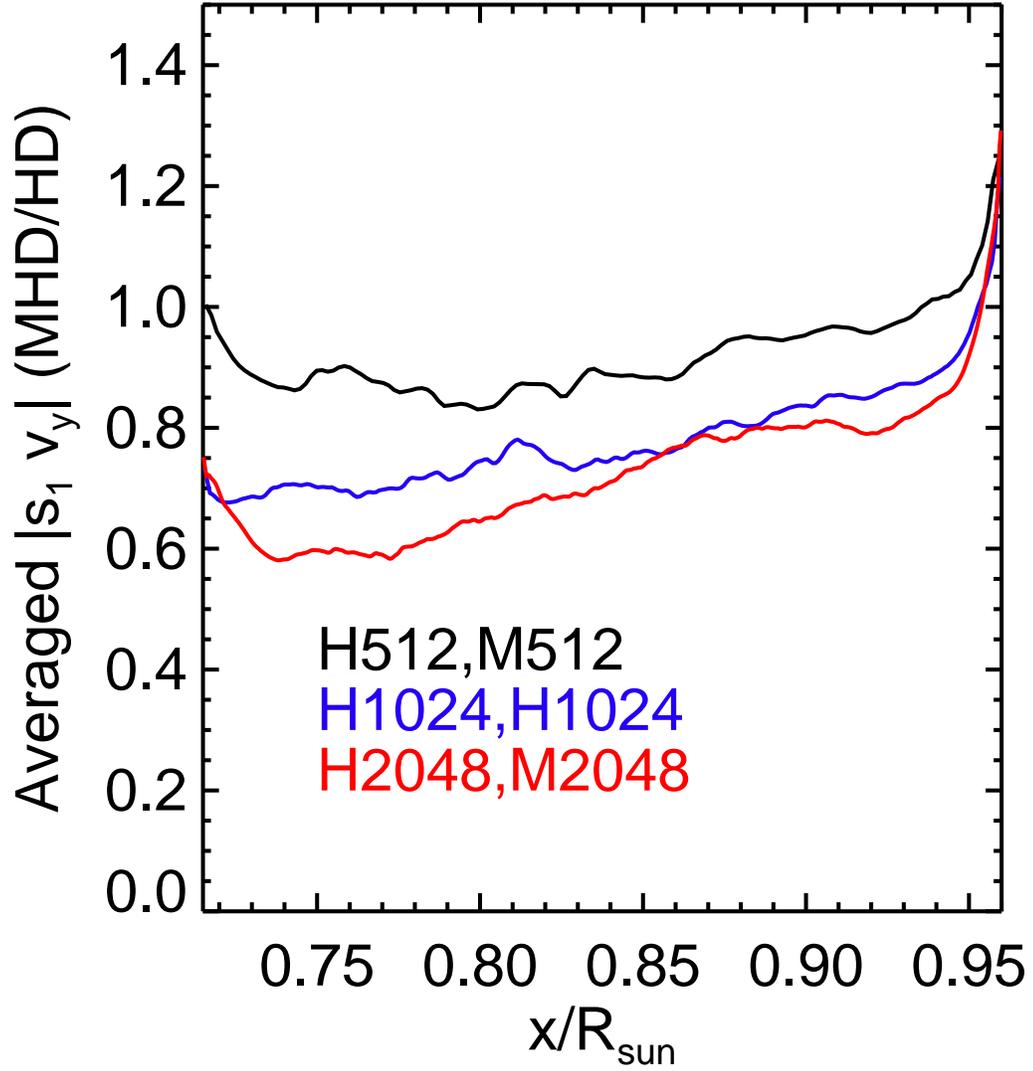}
 \caption{
 Ratio of horizontally averaged $|s_1v_y|$ between MHD and HD cases.
 The black, blue and red lines show the result in cases H(M)512,
 H(M)1024, and H(M)2048, respectively.
 \label{unsign_mixing}
 }
\end{figure}

Fig. \ref{HM2048_flux} shows the horizontally averaged vertical energy
fluxes. The definition of the enthalpy flux $F_\mathrm{e}$, the kinetic
flux $F_\mathrm{k}$, the radiative flux $F_\mathrm{r}$, the Poynting
flux $F_\mathrm{m}$, and total flux $F_\mathrm{t}$ are shown as
\begin{eqnarray}
 F_\mathrm{e} &=& \left(\rho_0 e_1 + p_1 - \frac{p_0\rho_1}{\rho_0}\right)v_r, \\
 F_\mathrm{k} &=& \frac{1}{2}\rho_0 v^2 v_r, \\
 F_\mathrm{r} &=& -\kappa_\mathrm{r}\rho_0 c_p \frac{dT_0}{dr} +
  \int_{x_\mathrm{min}}^x \Gamma(x)dx, \\
F_\mathrm{m} &=& \frac{c}{4\pi}(E_yB_z-E_zB_y),\\
 F_\mathrm{t} &=& F_\mathrm{e} + F_\mathrm{k} + F_\mathrm{r} + F_\mathrm{m},
\end{eqnarray}
where $e_1$ is the internal energy and ${\bf E}=-({\bf v}\times {\bf
B})/c$ is the electric field.
The artificial surface cooling is included in the radiative flux
$F_\mathrm{r}$ in order to avoid complexity in the figure.
We find a 
downward directed Poynting flux in our study that reaches 8\% of the
solar energy flux. This is three times larger than the value found in Paper
I. Whereas the reduction in the RMS velocity is large (25\%
reduction in the middle of the convection zone), the reduction of the
enthalpy flux is relatively small (12\% at maximum), which implies 
an increase of the entropy difference between up- and downflows. Fig. \ref{HM2048_pdf_ent}
shows PDFs of the entropy perturbation. The dotted and solid lines show
the results in cases H2048 and M2048, respectively. The black, blue and
red lines shows the results at $r=0.95R_\odot$, $0.8R_\odot$, and
$0.72R_\odot$, respectively. In every layer, the entropy perturbation,
especially in negative part, has larger amplitude in case
M2048. Figs. \ref{HM2048_2dpdf}a and b show joint PDFs with vertical
velocity $v_x$ and the entropy perturbation $s_1$ in cases H2048 and
M2048, respectively. In the MHD case (Fig. \ref{HM2048_2dpdf}b), the
larger amplitude of the entropy perturbation is achieved in the
downflow region. This improves the overall efficiency of the convective
energy transport.
Figs. \ref{HM2048se}a and b show the contour of the entropy at
$r=0.8R_\odot$. Unlike the vertical velocity (Figs. \ref{H2048vx}a and
\ref{M2048vxbx80}a), the presence of small-scale magnetic field leads to
smaller features in the entropy perturbation.
In addition the area of negative
entropy perturbation is reduced by 20\% at maximum. The area of negative
entropy perturbation is
30\% and 23\% at the middle of the convection zone in the cases H2048
and M2048, respectively.
This is also seen in the spectra of
entropy in Fig. \ref{HM2048_f_entropy}. In the deeper layer in particular
the small-scale entropy perturbation is amplified. As a result, the
enthalpy
flux in the middle of the convection is shifted to smaller scales
(Fig. \ref{HM2048_f_enthalpy}). In other words, strong magnetic
field enhances the small-scale energy transport in the middle
of the convection zone.
Fig. \ref{entropy_updown} shows the RMS values of the entropy in cases
H2048 and M2048. In the
downflow region (blue lines), the amplification of the entropy
perturbation is more
pronounced. The effect of the magnetic field in upflow region (red
lines) is mostly seen around the base of the convection zone.
Fig. \ref{ks_pdf2} shows the kurtosis and skewness of the entropy in
cases H2048 and M2048. The intermittency and asymmetry of the entropy
are increased by the magnetic field.
Figs. \ref{comp_rmsvel}c and d show that the horizontal
velocity is more suppressed by the magnetic field than the vertical
velocity. 
In addition, Fig. \ref{unsign_mixing} shows the ratio of
horizontally averaged $|s_1v_y|$ between MHD and HD cases. This
is a measure of the reduced horizontal mixing of internal energy
between up- and downflows.
We find a significant reduction
of the horizontal energy mixing with the presence of strong magnetic field.
This results in an increased and more efficient energy transport, i.e.
the enthalpy flux is not suppressed significantly, despite the reduction
of overturning mass.
\section{Summary}
We presented a series of high resolution MHD simulations in the solar
convection zone and studied small-scale dynamo action through the 
convection zone. For grid spacings
smaller than 350 km, the RMS magnetic field around the base of the
convection zone reaches 95\% of the equipartition field strength. 
Even around the top boundary 55\% of equipartition
magnetic field is maintained.  These values are consistent with those
required in \cite{2014ApJ...789..132R} to explain the observed field
strength of the quiet Sun. We find a significant Lorentz force feedback,
the convective RMS velocity is reduced to 50\% and 85\% compared to 
the hydrodynamic reference run at the top and bottom boundaries, respectively.
\par
The efficiency of the small-scale dynamo achieved in this study
is larger than that of the large-scale dynamo found in many recent global simulations
(i.e. we reach substantially stronger $B_\mathrm{rms}$ values in our setup with a 
small-scale dynamo alone).
\par
The significant suppression of the convective velocity does not lead to a
significant decrease of the enthalpy flux. This study shows that the
suppression of the velocity reduces the horizontal mixing of the
entropy between upflow and downflow regions and causes larger entropy
perturbations. This improves the efficiency of the
convective energy transport.
The reduction of the kinetic energy does not converge especially in the
lower convection zone where the convective scale is larger, which implies that 
Lorentz force feedback on convective flows is potentially larger than presented 
here. Even in that case the horizontal mixing of the entropy might be further reduced
leading to even stronger entropy perturbations between up- and downflows and more
efficient convective energy transport.  Overall we conclude that the Lorentz force feedback
from small-scale magnetic field is likely significant and has to be taken into account for
realistic simulations of the solar convection zone, which is currently not possible in
global simulations.  
\par
Recently \citet{2012PNAS..10911928H} presented rather stringent helioseismic constraints on 
convective velocities in the solar interior, which suggest that current convection simulations
overestimate convective velocities by up to several orders of magnitude. In a completely
independent study of solar supergranulation through numerical
simulations \citet{2014ApJ...793...24L}
found that high resolution simulations of the solar photosphere are consistent with observational
constraints in the uppermost 10~Mm of the convection zone, but overestimate convective
velocities by a factor of a few in more than 10~Mm depth. The work presented here indicates
that ubiquitous small-scale magnetic field throughout the solar convection zone might
contribute to some degree to resolving the issue, however, the reduction of convective velocities 
we find is still insufficient. Not completely unrelated to this is likely also the difficulty global
convection simulations have in reproducing a solar-like differential rotation in setups that transport
a full solar luminosity. The work by 
\citet{2014MNRAS.438L..76G,2014A&A...570A..43K}
suggests that a fast rotating equator requires
a lower Rossby number, i.e. lower convection velocities throughout the convection zone.
Recently \citet{2014ApJ...789...35F} showed that also Maxwell-stresses from a dynamo generated magnetic field
can significantly alter the sign and shape of differential rotation. To which degree this is due to large-scale
or small-scale magnetic field components remains at this point an open issue and requires  numerical 
simulations of efficient small-scale dynamos in a global  setup with rotation, which is work in progress.

\acknowledgements
We are grateful to Mark Miesch for helpful comments on the manuscript.
H. H. is supported by JSPS Postdoctoral Fellowship for Research Abroad.
The National Center for Atmospheric Research is sponsored by the
National Science Foundation. The results are obtained by using K
computer at the RIKEN Advanced Institute for Computational Science
(Proposal number hp140212) and the Fujitsu PRIMEHPC FX10 System
(Oakleaf-FX,Oakbrige-FX) in the Information Technology Center, The
University of Tokyo. This work was supported MEXT SPIRE and
JICFuS.

%\bibliographystyle{apj}
%\bibliography{apj-jour,reference}

\clearpage

\begin{table}
\begin{center}
\caption{Number of grid points and diffusivities in each
 case.\label{param}}
\begin{tabular}{lccc}
\tableline\tableline
Case & $N_x\times N_y\times N_z$ & $\kappa$, $\nu$, $\eta$
 $[\mathrm{cm^{2}\ s^{-1}}]$\\
\tableline
H256D, M256D & $72\times256\times256$ & $1\times 10^{12}$\\
H512, M512 & $144\times512\times512$ & -\\
H1024, M1024 & $288\times1024\times1024$ & -\\
H2048, M2048 & $576\times2048\times2048$ & -\\
\tableline
\tableline
\end{tabular}
\end{center}
\end{table}

\clearpage

\clearpage

\end{document}